\documentclass[journal]{vgtc}
 
\onlineid{1616}

\vgtccategory{Research}	

\vgtcpapertype{Data Transformations}

\title{TopoSZ: Preserving Topology in \texorpdfstring{\\}{} Error-Bounded Lossy Compression}

\author{
  \authororcid{Lin Yan}{0000-0001-7017-0329}, \authororcid{Xin Liang}{0000-0002-0630-1600}, \authororcid{Hanqi Guo}{0000-0001-7776-1834}, and \authororcid{Bei Wang}{0000-0002-9240-0700}
}

\authorfooter{
  \item
  	Lin Yan is with Argonne National Laboratory.
  	E-mail: lyan@anl.gov
  \item Xin Liang is with the University of Kentucky. E-mail: xliang@uky.edu.
  \item
  	Hanqi Guo is with the Ohio State University.
  	E-mail: guo.2154@osu.edu.
  \item Bei Wang is with the University of Utah.
  	E-mail: beiwang@sci.utah.edu
}

\abstract{Existing error-bounded lossy compression techniques control the pointwise error during compression to guarantee the integrity of the decompressed data. 
However, they typically do not explicitly preserve the topological features in data. 
When performing post hoc analysis with decompressed data using topological methods, 
preserving topology in the compression process to obtain topologically consistent and correct scientific insights is desirable. 
In this paper, we introduce TopoSZ, an error-bounded lossy compression method that preserves the topological features in 2D and 3D scalar fields. 
Specifically, we aim to preserve the types and locations of local extrema as well as the level set relations among critical points captured by contour trees in the decompressed data. 
The main idea is to derive topological constraints from contour-tree-induced segmentation from the data domain, and incorporate such constraints with a customized error-controlled quantization strategy from the SZ compressor (version 1.4).
Our method allows users to control the pointwise error and the loss of topological features during the compression process with a global error bound and a persistence threshold. 
}

\keywords{Lossy compression, contour tree, topology preservation, topological data analysis, topology in visualization}

\graphicspath{{figs/}} 

\usepackage{tabu}
\usepackage{booktabs}
\usepackage{lipsum}
\usepackage{mwe}                       

\usepackage{amssymb}
\usepackage{amsmath} 
\usepackage{amsfonts}

\usepackage{multirow}
\usepackage{booktabs,lipsum}
\usepackage{mathtools}
\usepackage{textcomp}

\newcommand{\para}[1]        {\vspace{1pt}\noindent{\textbf{#1}}}
\newcommand{\denselist}{\vspace{-3pt} \itemsep -2pt\parsep=-1pt\partopsep -2pt}

\newcommand{\VF}        {\mm{\mathsf{Viscous\,Fingers}}}
\newcommand{\IB}        {\mm{\mathsf{Isabel}}}
\newcommand{\NYX}    {\mm{\mathsf{NYX}}}
\newcommand{\TN}     {\mm{\mathsf{Tornado}}}
\newcommand{\EW}        {\mm{\mathsf{E3SM\,Wind}}}
\newcommand{\HF}        {\mm{\mathsf{Heated\,Flow}}}
\newcommand{\AF}        {\mm{\mathsf{Tangaroa}}}

\newcommand {\mm}[1] {\ifmmode{#1}\else{\mbox{\(#1\)}}\fi}

\newcommand{\Xspace}        {\mm{\mathbb{X}}}
\newcommand{\Rspace}        {\mm{\mathbb{R}}}

\newcommand{\lb}  {\mm{\mathcal{L}}}
\newcommand{\ub} {\mm{\mathcal{U}}}
\newcommand{\lbi}        {\mm{l}}
\newcommand{\ubi}      {\mm{u}}
\newcommand{\vareps}      {\mm{\varepsilon}}

\newcommand{\etal}{{et al.}}

\newcommand{\ie}{{i.e.}}
\newcommand{\wrt}{{w.r.t.}}
\newcommand{\cf}{{c.f.}}

\newcommand{\tool}        {\mm{\text{TopoSZ}}}
\newcommand{\myedit}[1]{{\textcolor{black}{#1}}}

\begin{document}

\firstsection{Introduction}
\maketitle
\firstsection{Introduction}
\label{sec:introduction}
Advances in high-performance computing allow researchers to generate extremely large volumes of scientific data. 
Such data pose significant challenges for memory resources and transmission bandwidths in scientific analysis and visualization.
\emph{Lossy compression}, a class of data compression techniques that involves some loss of information via inexact approximations and partial data discarding, has the potential to address these challenges as it enables a significant reduction in data size. 
Lossy compression has been widely used in computer graphics (e.g.,~\cite{AaidPearlman1996}), scientific visualization (e.g.,~\cite{Chow1997,SchneiderWestermann2003}), storage systems, databases, and software applications; see~\cite{CappelloDiLi2019,JayasankarThirumalPonnurangam2021} for surveys. 
 
To preserve data integrity, many scientific applications require \emph{error-bounded} lossy compression to control the extent to which data are altered during  compression.  
For instance, compressors such as SZ~\cite{TaoDiChen2017,liangDiTao2018}, ZFP~\cite{Lindstrom2014}, and FPZIP~\cite{LindstromIsenburg2006} guarantee the pointwise error between the original and the decompressed data.
However, a constraint on pointwise error alone may alter  the features of interest from the original data. 
Inconsistencies between features extracted from the original and the decompressed data may lead to inconsistent or incorrect scientific insights during the post hoc analysis of the decompressed data. 

Topological descriptors for scalar field data, such as merge trees~\cite{BeketayevYeliussizovMorozov2014}, contour trees~\cite{CarrSnoeyinkAxen2003}, Reeb graphs~\cite{Reeb1946}, and Morse and Morse-Smale complexes~\cite{GerberPotter2012, EdelsbrunnerHarerZomorodian2003, EdelsbrunnerHarerNatarajan2003}, have been widely used in scientific data analysis and visualization. 
Topological methods based on these descriptors have found applications in science and engineering, such as symmetry detection in materials science~\cite{ThomasNatarajan2011, SaikiaSeidelWeinkauf2014, ThomasNatarajan2014, SaikiaSeidelWeinkauf2015, SridharamurthyMasoodKamakshidasan2020} and cloud tracking~\cite{DoraiswamyNatarajanNanjundiah2013, EngelkeMasoodBeren2020, NilssonEngelkeFriederici2020} in climate science; see~\cite{YanMasoodSridharamurthy2021} for a survey. 
\myedit{Preserving topological descriptors is important if data compression is part of the analysis or storage pipeline within these applications. 
For example, Yan \etal~\cite{YanMasoodRasheed2022} utilized merge trees to summarize the distribution of ocean eddies. 
Missed or added topological features during data compression may lead to misleading observations of the climate system.}

In this paper, we introduce {\tool}, an error-bounded lossy compression technique that preserves topological features in 2D and 3D scalar field data. 
We use the \emph{contour tree}, a topological descriptor that records the connectivity among level sets of a scalar field, to aid our compression process and evaluate its effectiveness. 
When performing post hoc analysis with decompressed data using topological methods, {\tool} guarantees topological consistencies between the original and the decompressed data.  
Specifically, we aim to preserve the types and locations of local extrema as well as the level set relations among critical points captured by contour trees in the decompressed data. 
The main idea is to derive topological constraints from contour-tree-induced segmentation from the data domain, and incorporate such constraints with a customized error-controlled quantization strategy from the SZ compressor \myedit{(version 1.4~\cite{TaoDiChen2017}, referred to as SZ-1.4 in the rest of the paper)}.

\begin{figure}[t]
    \centering
    \includegraphics[width=0.8\columnwidth]{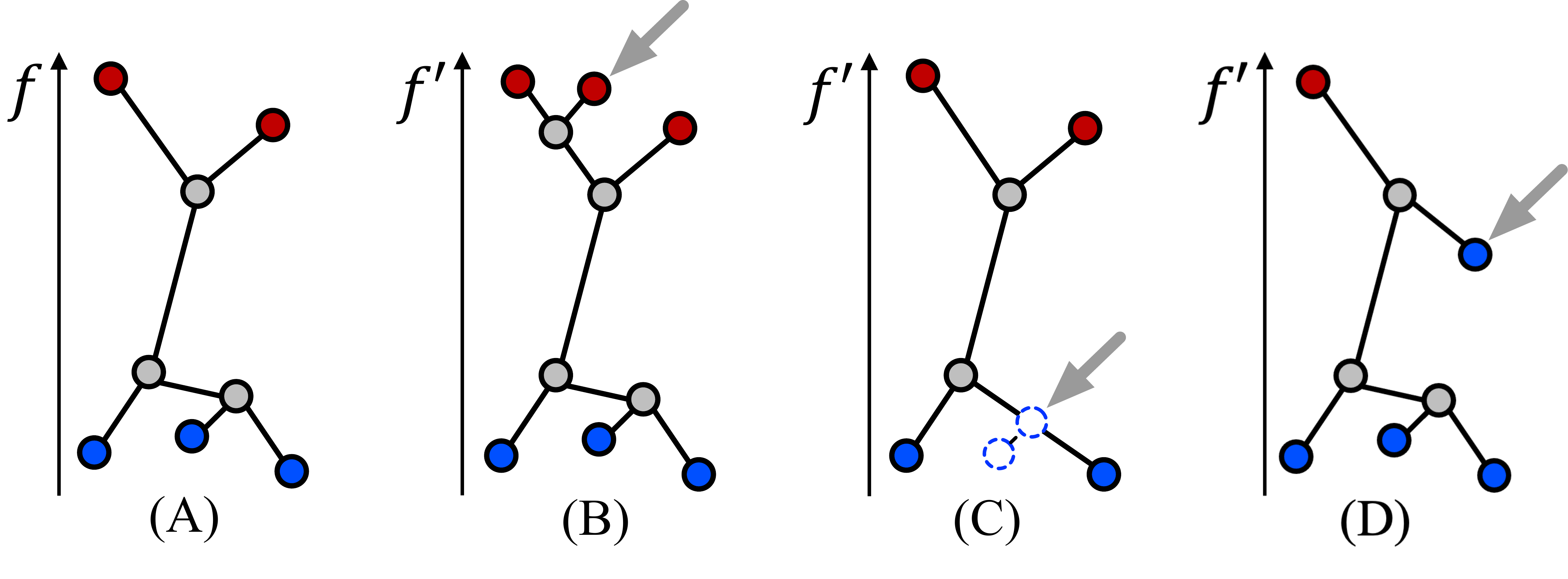}
    \vspace{-4mm}
    \caption{Examples of false cases (pointed by arrows) in preserving the contour trees during compression. (A) The contour tree that arises from the original scalar field $f$. (B-D) Contour trees that arise from the decompressed scalar field $f'$: (B) false positive, (C) false negative, and (D) false type.} 
    \vspace{-4mm}
    \label{fig:FPFNFT}
\end{figure}

To quantify the level of topology preservation, we introduce three types of false cases in preserving the topology captured by contour trees, which are inspired by the work of Liang \etal~\cite{LiangGuoDi2020}: false positives (FPs), false negatives (FNs), and false types (FTs). 
A false positive occurs when a new branch appears in the contour tree of the decompressed data, which does not exist in the same position of the contour tree from the original data; \cf~\cref{fig:FPFNFT}(A) and (B). 
A false negative occurs when a branch of the contour tree from the original data is missing from the contour tree of the decompressed data; \cf~\cref{fig:FPFNFT}(A) and (C). 
A false type occurs when the corresponding critical points do not match in type between the contour trees of the original and the decompressed data; \cf~\cref{fig:FPFNFT}(A) and (D). 
Given a persistent simplification threshold $\varepsilon$ that controls the importance of features to be preserved, our method  preserves a simplified contour tree from the original data without any FPs, FNs, and FTs. In this paper, we aim to preserve topological features and retain pointwise error control in the lossy compression of 2D and 3D scalar fields. Our contributions include: 
\begin{itemize} \denselist
\item \myedit{A contour-tree-based iterative strategy to derive topological constraints as lower and upper bounds for lossy compression;}
\item A customized error-controlled quantization strategy to incorporate topological constraints into the data compression process;
\item Comparison with the state-of-the-art  error-bounded lossy compressors in terms of compression ratio, PSNR (peak signal-to-noise ratio), and topological metrics, using a number of scientific datasets.
\end{itemize}
\section{Related Work}
\label{sec:related-work}

\para{Error-bounded lossy compression.}
Data compression techniques can be classified based on information loss during data reduction, namely, lossless and lossy compression. Lossless compressors, such as~\cite{HowardVitter1992, BurtscherRatanaworabhan2007, BurtscherRatanaworabhan2008, TiernyPascucci2012}, ensure   that the decompressed (reconstructed) data inherit all properties of the original data; however, these techniques often suffer from  insufficient compression rates in practice. 
Lossy compressors usually achieve a higher data reduction ratio than lossless compressors with an acceptable amount of information loss. 
\myedit{For example, lossless compression ratios range from about 1.5 to 3 for image compression, whereas lossy compression techniques offer compression ratios in excess of 20 with virtually no loss in visual fidelity.}
Lossy compressors can be further categorized into non-error-bounded and error-bounded ones.
This paper focuses on error-bounded lossy compressors since they usually deliver higher compression ratios than the lossless ones and are more precise than the non-error-bounded lossy compressors.

Error-bounded lossy compression may be classified as truncation-based, prediction-based, and transformation-based methods.
Truncation-based error-bounded lossy compressors include the work of Gong~\etal~\cite{GongRogersJenkins2012}. 
The authors designed MLOC for large scientific analysis by truncating double-precision floating-point numbers into multiple precision levels, which are further used to support data access with certain resolution requirements. 
An example of a prediction-based method is ISABELA~\cite{LakshminarasimhanShahEthier2011}, which transfers data points to smooth curves using B-splines for prediction tasks. 
Another prediction-based method is FPZIP~\cite{LindstromIsenburg2006}, which uses the Lorenzo predictor~\cite{IbarriaLindstromossignac2003} to estimate the value on an unknown vertex based on known cube vertices. 
Both the predicted and the actual values are mapped to an integer representation to avoid underflow, together with arithmetic coding for residuals. 
SZ~\cite{liangDiTao2018,LiangZhaoDi2022,TaoDiChen2017,ZhaoDiDmitriev2021,ZhaoDiLiang2020} is a family of prediction-based compressors featuring on an adaptive selection of the best-fit prediction methods (e.g., Lorenzo predictor, regression-based predictors, and interpolation-based predictors).
 Instead of using arithmetic coding for residuals as FPZIP does, SZ performs a linear-scaling quantization~\cite{TaoDiChen2017}  to convert the difference between the predicted value and the original value for each datapoint to an integer. 
These quantization integers are encoded by customized Huffman coding and can be further compressed by a lossless compressor such as ZSTD~\cite{ZSTD2021} and GZIP~\cite{GZIP2021}. 
ZFP~\cite{Lindstrom2014} is a transformation-based error-bounded compressor. It uses a custom orthogonal block transform to de-correlate the values of blocks and encodes the transform coefficients for compression.

In this paper, we customize the SZ-1.4 compressor to preserve topology, since several prior studies~\cite{DiTaoLiang2018, LiangDiLi2019,ZhaoDiLiang2020} have verified that SZ yields the best compression quality under the same error bound among all the prediction-based compressors. 
However, in general, any prediction-based compressor can be customized for topology-preserving compression using our framework. 

\para{Topology-preserving compression.}
To the best of our knowledge, Soler~\etal~\cite{SolerPlainchaultConche2018} presented the first and only compression technique for scalar field data with some topological guarantees. 
Their technique, referred to as TopoQZ in this paper, is developed based on a topologically adaptive quantization of the data range. 
In particular, TopoQZ supports a controlled loss of topological features. That is, given a persistent threshold $\varepsilon$ that quantifies the target feature size, TopoQZ ``preserves the critical point pairs with persistence greater than $\varepsilon$ and destroys all pairs with a smaller persistence.''~\cite{SolerPlainchaultConche2018}  

{\tool} inherits the compression pipeline in SZ (SZ-1.4~\cite{TaoDiChen2017}) and modifies the quantization process (see~\cref{sec:quantization} for details).  
Thus, our framework differs significantly from TopoQZ~\cite{SolerPlainchaultConche2018}.
First, TopoQZ does not enforce pointwise error control. {\tool} controls pointwise error explicitly, thus ensuring the pointwise compression quality besides topology preservation. 
Second, TopoQZ uses a quantization of the scalar field, whereas {\tool} uses a customized error-controlled quantization encoder to reduce data size significantly during the compression process. 
We give a detailed comparison between {\tool} and TopoQZ in terms of  compression qualities and compression ratios in~\cref{sec:results}. 

\para{Contour trees.}
We use contour trees, which capture the connectivity among level sets of scalar fields, to design and evaluate our topology-preserving compression technique. 
Contour trees have been used to detect symmetry in bimolecular structures~\cite{ThomasNatarajan2011}, identify and track the oxy-deoxy process in hemoglobin dynamics~\cite{SohnBajaj2006}, and design shape matching algorithms for electrostatic potentials~\cite{ZhangBajajBaker2004}. Contour trees are also used in studying scalar field ensembles, such as their distributions~\cite{ZhangBajajBaker2004}, summarization~\cite{LohfinkWetzelsLukasczyk2020}, and uncertainty visualization~\cite{WuZhang2013}.

\para{Mesh simplification} \myedit{is another effective way to reduce data for analysis and storage. Chiang and Lu~\cite{ChiangLu2003} introduced a mesh simplification technique that preserves isosurfaces with geometric error control. However, we consider mesh simplification and compression to be fundamentally different approaches in data reduction. Therefore, we do not compare against mesh simplification in this paper.}

\section{Technical Background}
\label{sec:background}
Our approach has four technical gradients: SZ-1.4, contour tree, persistence simplification, and contour-tree-induced segmentation. 

\subsection{\myedit{SZ-1.4 Compressor}}
\label{sec:classicSZ}

We review the SZ-1.4 compressor and customize it in~\cref{sec:quantization} to build {\tool}.  
A newer version of the SZ compressor (referred to as SZ3~\cite{LiangZhaoDi2022,ZhaoDiDmitriev2021}) uses an interpolation algorithm to improve the compression ratios when the error bound is relatively high, and switches to the Lorenzo predictor when the error bound becomes low. 
Since TopoSZ usually targets the high-precision (low-error-bound) cases, we directly use the Lorenzo predictor in our implementation, which is part of SZ-1.4.

\para{An overview of SZ-1.4.}
SZ-1.4 is a prediction-based lossy compressor designed for scientific data that strictly controls the pointwise compression error. 
That is, a global error bound $\xi$ guarantees $|f(x)-f'(x)|<\xi$, where $f(x)$ and $f'(x)$ are the original and the decompressed values of any datapoint $x$ in the domain. 
The implementation of SZ-1.4 involves four steps: data prediction, linear-scaling quantization, customized variable-length encoding, and lossless compression. 

\para{\underline{Step 1:~data prediction.}} SZ-1.4 predicts the value of each datapoint with a \emph{Lorenzo predictor}~\cite{IbarriaLindstromossignac2003} based on the decompressed values of its neighbors. 

\para{\underline{Step 2:~linear-scaling quantization.}} SZ uses linear quantization on the difference between the original and the predicted values for each datapoint.
In SZ-1.4~\cite{TaoDiChen2017}, the difference is converted to an integer based on the global error bound $\xi$. 

\para{\underline{Step 3:~customized variable-length coding.}} SZ-1.4 encodes the quantization integers from the previous step by a customized Huffman encoding~\cite{Abu-MostafaMcEliece2000} to reduce the data size for storage. 

\para{\underline{Step 4: lossless compression.}} In the last step, SZ-1.4 further compresses the unpredictable data from step 2 and the encoded quantization from step 3 by a lossless compressor ZSTD~\cite{ZSTD2021}. 

\begin{figure}[!ht]
    \centering
    \vspace{-2mm}
    \includegraphics[width=1.0\columnwidth]{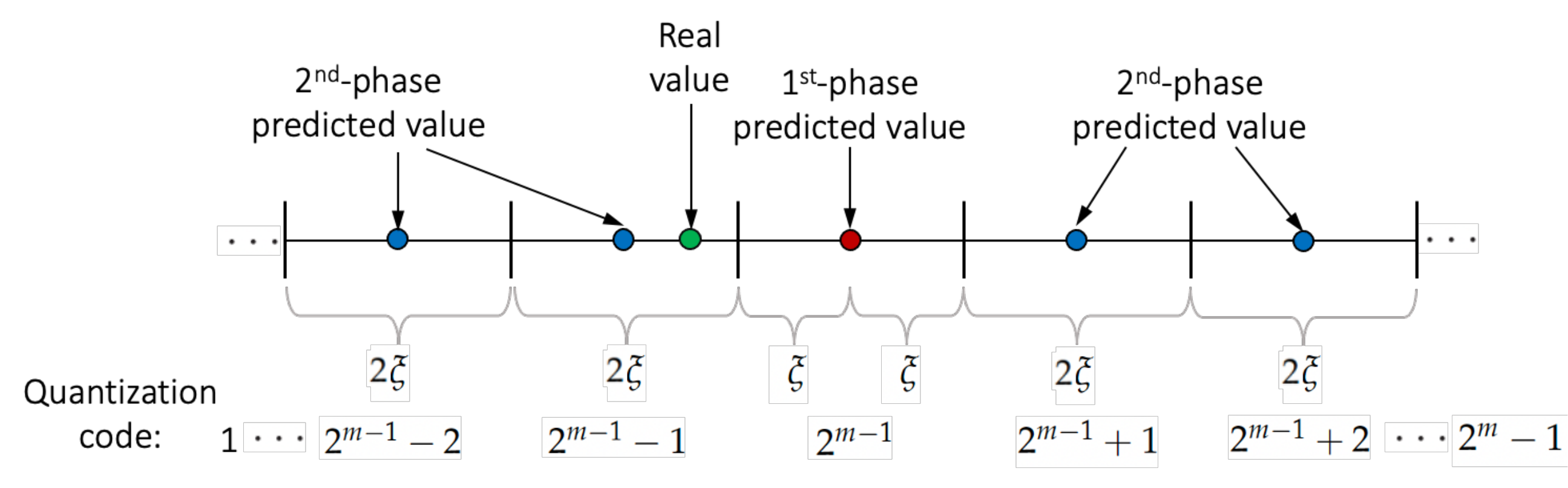}
    \vspace{-8mm}
    \caption{The design of an error-controlled quantization encoder in the SZ-1.4 based on a linear scaling of the error bounds. Image reproduced from~\cite[Fig.~2]{TaoDiChen2017}.} 
    \label{fig:error-control-sz}
     \vspace{-2mm}
\end{figure}

\para{Linear-Scaling quantization.}	
The key differences between {\tool} and SZ-1.4 are that {\tool} derives the lower/upper bounds for each datapoint, and it revises the quantization method of SZ-1.4 (referred to as step 2) to guide the quantization process for topology preservation. 
Therefore, we review an SZ-1.4 implemented by Tao {\etal} that introduces the notion of an error-controlled \emph{quantization encoder}; see~\cite{TaoDiChen2017} for details. 
Let $\xi$ denote the global error bound. The design of the SZ-1.4 quantization encoder is shown in \cref{fig:error-control-sz}. 
Let $m$ denote the number of bits needed to encode $2^m-1$ number of quantization codes. We set $m=16$. 

For a datapoint $x \in \Xspace$, we first compute its \emph{1st-phase predicted value} by using a prediction model. 
For example, Tao {\etal} used a multilayer prediction model (e.g., a Lorenzo predictor is a special case of the multilayer predictor).  
Such a value is presented by a red point in~\cref{fig:error-control-sz}. 
Then, we generate a number of \emph{2nd-phase predicted values}  by scaling the error bound linearly. These values are represented by the blue points in \cref{fig:error-control-sz}.  
The distance between any two adjacent 2nd-phase predicted values equals $2\xi$. 
By construction, each (1st-phase or 2nd-phase) predicted value lies in the middle of an interval with the length of $2\xi$, and these intervals do not overlap.
In total, there are $2^m-1$ intervals indexed by $\{1, \dots, 2^m-1\}$.
The 1st-phase predicted value (red point) falls in the interval indexed by $2^{m-1}$.  

In SZ-1.4, when the real value of the datapoint $f(x)$ (green point) falls into a certain interval, we mark it as being \emph{predictable} and use its corresponding predicted value from the same interval to represent its value $f'(x)$ in the compression. By construction, $|f(x) - f'(x)| \leq \xi$.  
However, if $f(x)$ does not fall into any interval, we mark the datapoint as being \emph{unpredictable}.

Since there are $2^m-1$ intervals, each predicted value can be encoded by the index of its corresponding interval. 
Then ``we use the codes of $1, \dots, 2^m-1$ to encode predictable data  and use the code of $0$ to encode unpredictable data''~\cite{TaoDiChen2017}. 
This process is referred to as the quantization encoding. 

\subsection{Contour Tree and Persistence Simplification}
Given a scalar field $f: \Xspace \to \Rspace$ defined on a simply connected domain $\Xspace$, a \emph{contour} is a connected component of the \emph{level set} $\Xspace_t=f^{-1}(t)$ of $f$, for some $t \in \Rspace$.
A contour tree captures the connectivity among the contours of $f$. 
Two points $x,y \in \Xspace$ are \emph{equivalent}, denoted as $x \sim y$, if $f(x)=f(y)=t$, and $x$ and $y$ belong to the same contour of $f$. The \emph{contour tree}, $T(\Xspace, f) = \Xspace/{\sim}$, is the quotient space obtained by identifying equivalent points.  
For a point $x \in \Xspace$, all points $y \in \Xspace$ that are equivalent to $x$ forms its \emph{equivalence class}. 

\begin{figure}[!ht]
    \centering
    \includegraphics[width=0.8\columnwidth]{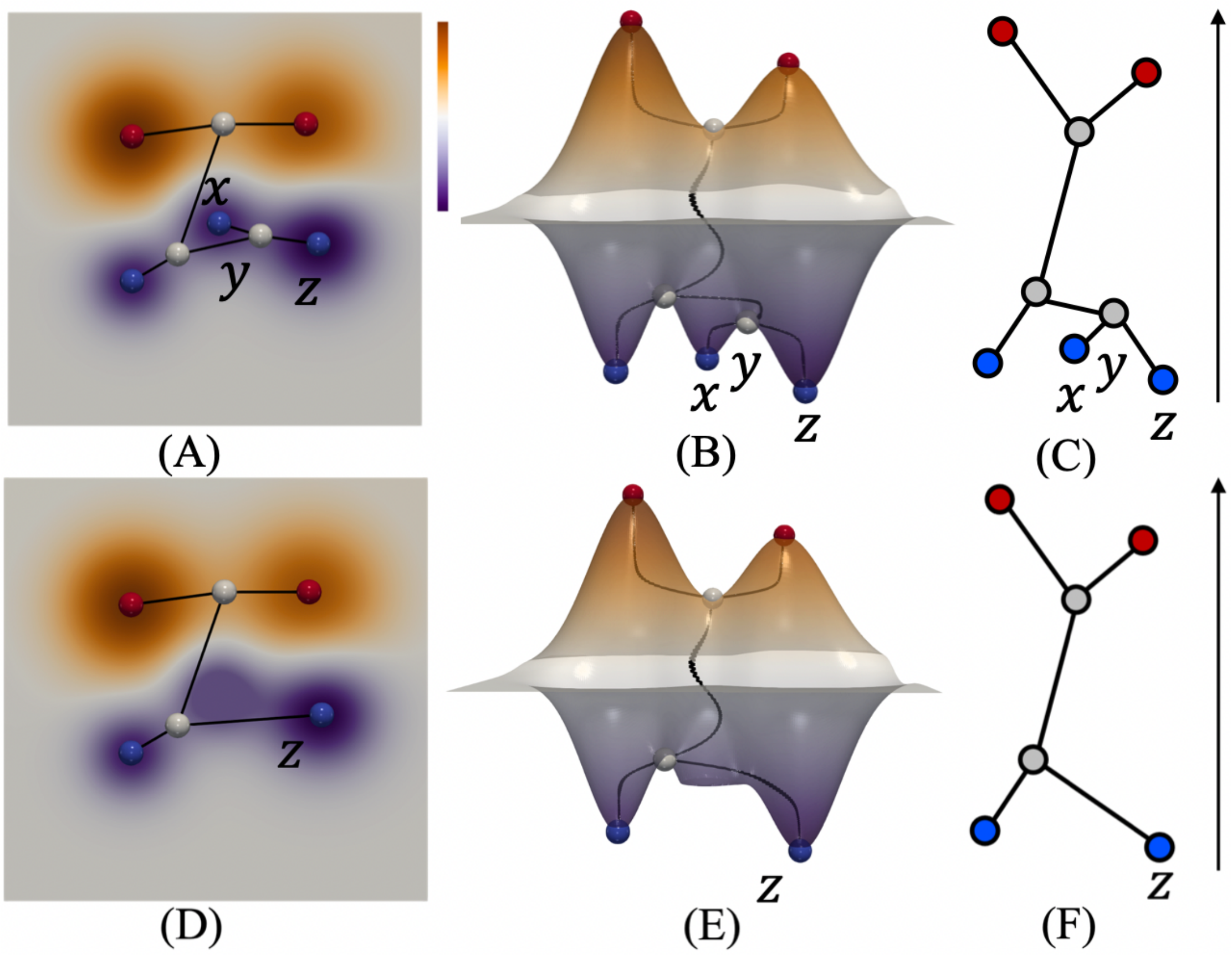}
    \vspace{-2mm}
    \caption{A contour tree of a 2D scalar field $f$ generated by a mixture of Gaussian functions. (A) The scalar field visualized  with an embedded contour tree $T$. (B) The graph of $f$---the set of all ordered pairs $(x, f (x))$---is visualized together with the corresponding contour tree of $f$. (C) Abstract (straight-line) visualization of the contour tree $T$ where nodes are equipped with scalar field values. (D)-(F) The minimum-saddle branch $(x,y)$ of $T$ is simplified.} 
    \label{fig:ct}
    \vspace{-2mm}
\end{figure}

As $t$ increases in the range of $f$, a contour tree captures the evolution of the contours of $f$; see~\cite[Sect. VI.4]{EdelsbrunnerHarer2010}  for a formal treatment.
Intuitively, the contour tree of $f$ is the graph obtained by a continuous contraction of each contour of $f$ to a single point. Nodes in the contour tree have a one-to-one correspondence with the \emph{critical points} of $f$, at which the topology of the contours changes. 
Assume $f$ is a Morse function, and critical points are locations where the derivative of $f$ is zero. 
For a 2D scalar field, critical points are local maxima, local minima, and saddles. 
\cref{fig:ct} gives an example. Blue nodes represent the creation of a component at a local minimum, gray nodes (saddles) represent the merging or splitting of components, and red nodes represent the disappearance of a component at a local maximum; see a straight-line drawing in \cref{fig:ct}(C). 
We display a scalar field with its corresponding contour tree embedded in the scalar field and the graph of the scalar field; see \cref{fig:ct}(A) and (B), respectively.  
For a 3D scalar field, critical points are local maxima, local minima, and saddles of order 1 and 2.  
Local maxima and local minima are referred to as local extrema. 
An \emph{edge} in a contour tree is a line joining a pair of critical points. 

A contour tree of real-world data typically contains a number of tiny branches due to the noise in the data.
A persistence-based~\cite{CarrSnoeyinkVan2010,GueunetFortinJomier2016,AthawaleMaljovecYan2020} simplification strategy may be used to simplify a contour tree, which also leads to the simplification of the scalar field~\cite{PascucciColeMcLaughlin2004, TakahashiTakeshimaFujishiro2004, CarrSnoeyinkVan2010}.
\cref{fig:ct} (E-F) illustrates the persistence simplification of a contour tree.

\subsection{Contour Tree Induced Segmentation}
\label{sec:segmentation}

\begin{figure}[!ht]
    \centering
    \vspace{-3mm}
    \includegraphics[width=\columnwidth]{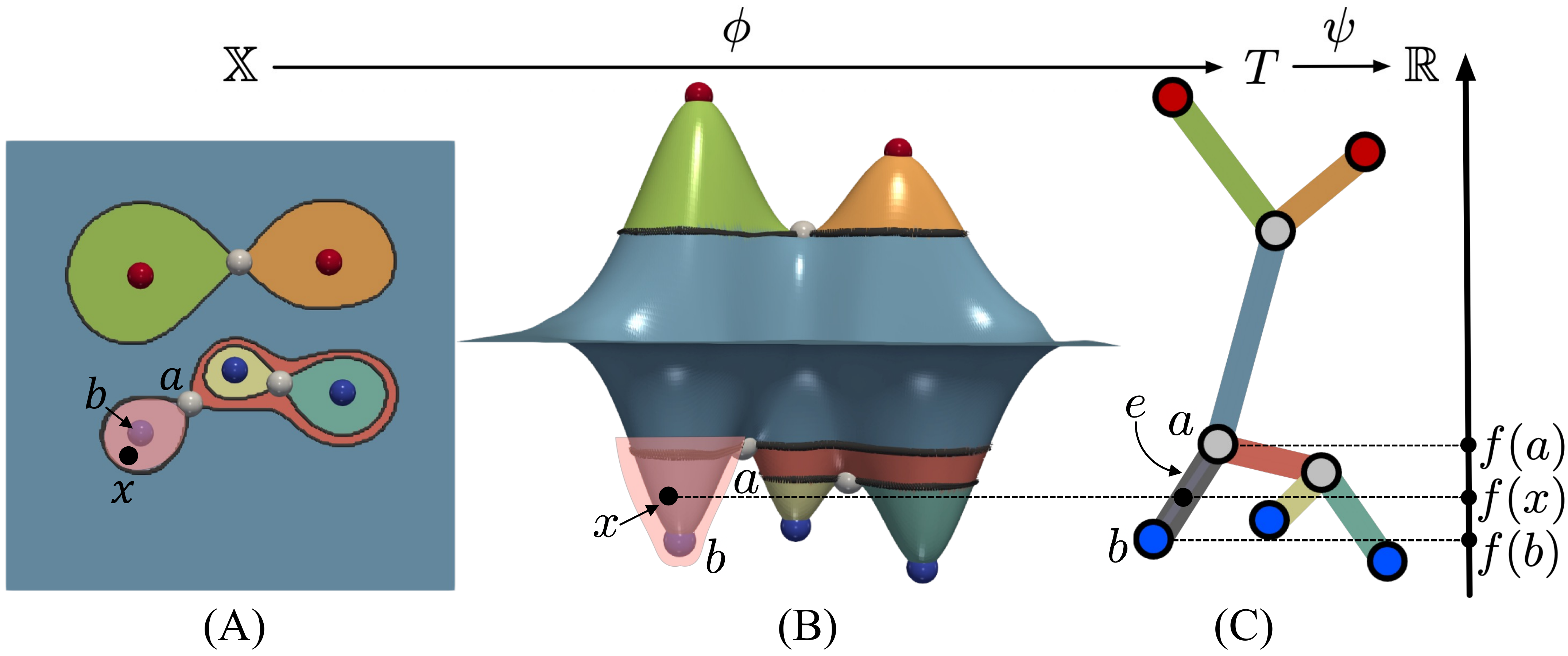}
    \vspace{-6mm}
    \caption{\myedit{A segmentation induced by the contour tree in \cref{fig:ct}. (A) $f$, (B) the graph of $f$, and (C) the contour tree $T$. 
     Edges of the contour tree in (C) and their pre-images  (\ie,~the topological regions) in (A) and (B) are shown with the same color.}} 
    \label{fig:ct-segmentation}
    \vspace{-2mm}
\end{figure}

By construction, a contour of a scalar field $f: \Xspace \to \Rspace$ maps to a point in the contour tree $T:=T(\Xspace, f)$. 
$f$ can be decomposed into $f =\psi \circ \phi$, where $\phi: \Xspace \rightarrow T$ maps each point in $\Xspace$ to its equivalence class in $T$ and $\psi: T \rightarrow \Rspace$ maps each point in $T$ to its $f$ value (see \cref{fig:ct-segmentation}). 

The pre-image of $\phi$ at critical points induces a partition of $\Xspace$. 
By construction, the pre-image $\phi^{-1}(e)$ of an edge $e \in T$ is connected and its function values vary monotonically.  
The pre-images of all edges in the contour tree thus give rise to a partition of $\Xspace$, referred to as the \emph{contour tree induced segmentation}, and each region in the partition is referred to as a \emph{topological region} (see~\cref{fig:ct-segmentation}A).  
A topological region is monotonic and contains only regular points in its interior. 
Furthermore, we can apply persistence simplification of the contour tree $T$ using branch decomposition, which is equivalent to iteratively removing its shortest (by height) leaf-connecting edges.

\begin{figure}[!ht]
    \centering
    \includegraphics[width=1.0\columnwidth]{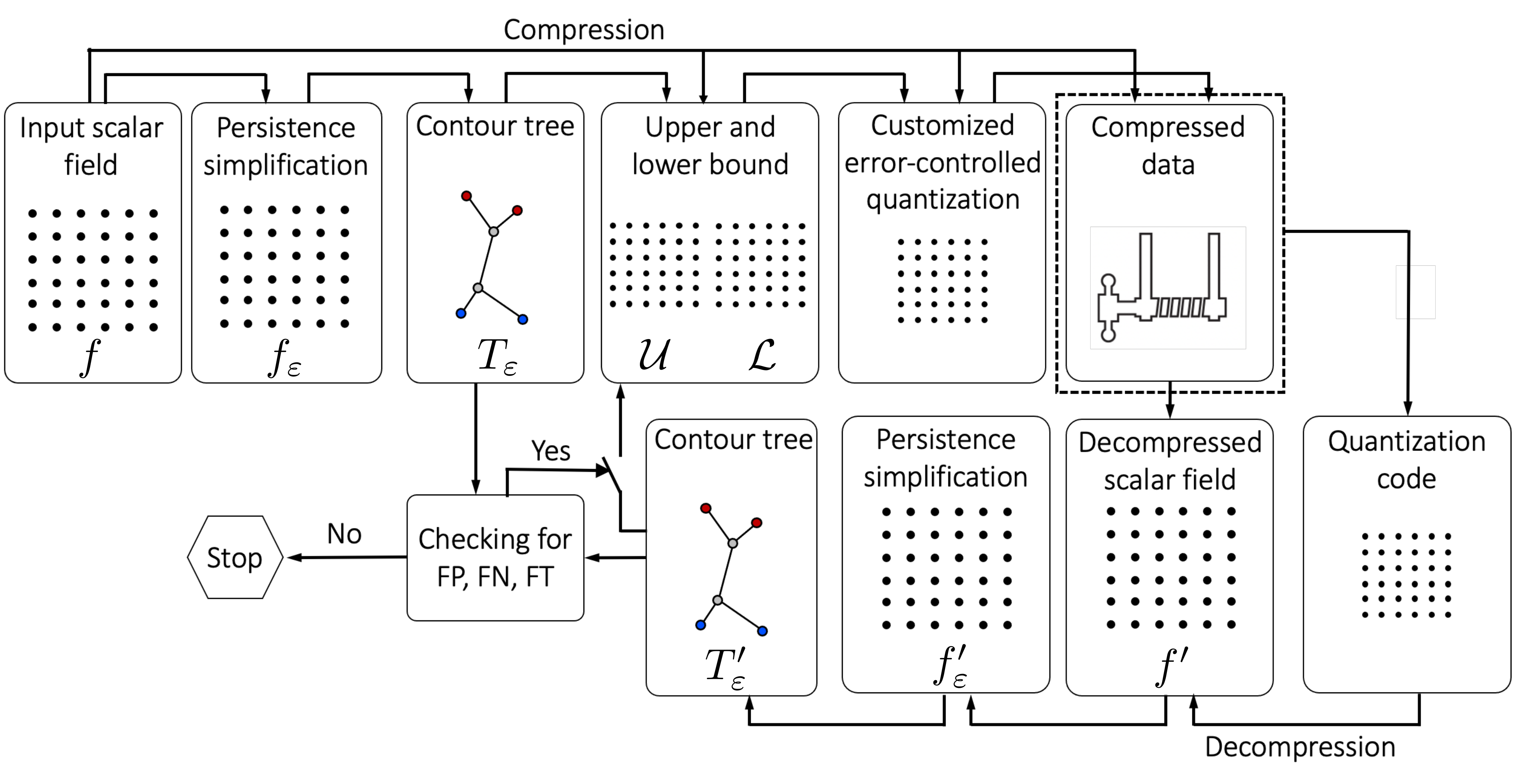}
    \vspace{-6mm}
    \caption{TopoSZ pipeline for topology-preserving compression. } 
    \label{fig:pipeline}
    \vspace{-6mm}
\end{figure}

\section{Method}
\label{sec:method}
In this section, we propose a novel lossy compressor---TopoSZ---that preserves the topology of data.  
TopoSZ modifies the SZ compressor (version 1.4) by explicitly imposing pointwise upper and lower error bounds that encode topological constraints derived from contour trees. 

\para{Notations.}
Let $\xi$ and $\vareps$ be the user-defined global error bound and the persistence threshold, respectively.  
Let $f: \Xspace \to \Rspace$ denote the original scalar field and $f_{\vareps}$ its $\vareps$-simplification. 
Similarly, let $f'$ denote the decompressed scalar field and $f'_{\vareps}$ its $\vareps$-simplification. 
Let $T_{\vareps}$ and $T'_{\vareps}$ be the contour trees of $f_{\vareps}$ and $f'_{\vareps}$, respectively.  
In practice, we assume $f$ is defined on $\Xspace := \{x_1, \dots,x_N\}$, a finite set of sampled datapoints. 
The input to TopoSZ includes $f$, $\xi$, and $\vareps$. 

\para{Overview.} 
TopoSZ compresses the original scalar field $f$ such that the decompressed scalar field $f'$ contains the same set of local extrema and gives rise to the same contour tree up to persistent simplification. 
In other words, TopoSZ ensures that $f_{\vareps}$ and $f'_{\vareps}$ contain the same set of local extrema, and $T_{\vareps} = T'_{\vareps}$, {\ie}, $T'_{\vareps}$ is free from false cases. 

An overview of our pipeline is shown in~\cref{fig:pipeline}. 
First, we impose topological constraints by computing the upper and lower error bounds $\ub := (\ubi_1, ..., \ubi_N)$ and $\lb := (\lbi_1, ..., \lbi_N)$ for all datapoints, using the topological regions defined by the contour-tree-induced segmentation (\cref{sec:segmentation}). 
These upper and lower bounds quantify the range of functional perturbations that do not change the level set relations and thus maintain topology during compression.

During the compression process, TopoSZ takes $\ub$, $\lb$ and $\xi$ as input. 
We compare the contour tree $T'_\vareps$ after decompression with the original contour tree $T_\vareps$. 
If no false cases occur, TopoSZ stops the compression process. 
Otherwise, $f'_{\vareps}$ and its contour tree $T_\vareps'$ are used as input for another round of compression with updated upper and lower bounds for a subset of datapoints until all false cases are eliminated from the decompressed data; see \cref{sec:ublb} for details. 

Second, we propose in \cref{sec:quantization} a novel error-controlled quantization encoder for TopoSZ by modifying the one from the SZ-1.4.  
Using such a customization, TopoSZ takes as input pointwise upper and lower error bounds $\ub$ and $\lb$ together with a global error bound $\xi$, and controls the compression error for every datapoint.

\subsection{Computing Contour-Tree-Based Error Bounds}
\label{sec:ublb}

To preserve the contour tree during compression, we need to preserve the locations and types of local extrema by eliminating false positives (FPs), false negatives (FNs), and false types (FTs). 
The key idea is to use contour-tree-induced segmentation (\cref{sec:segmentation}) to provide fine-grained controls of error bounds in topological regions. 
In particular, we preserve the critical points and the monotonicity of the scalar field in topological regions up to persistence simplification. 
Computing the upper and lower bounds for TopoSZ involves two key steps: initialization and iteration. 
Iteration may not be needed if the initialization step eliminate all false cases.

\subsubsection{Initialization} 
\label{sec:initialization}
As illustrated in~\cref{fig:ct-segmentation}, given an edge $e$ of the contour tree $T$, let $a, b \in \Xspace$ denote the critical points of $f$ in $\Xspace$ that correspond to the endpoints of $e$. Without loss of generality, we suppose $f(a)>f(b)$. 
For a point $x \in \Xspace$ in the pre-image $\phi^{-1}(e)$, its lower bound $\lbi$ and upper bound $\ubi$ are defined as
\begin{equation}
\left \{
  \begin{aligned}
    &\lbi=f(b), \ubi = f(a), && \forall x\in \phi^{-1}(e), x\neq a, x \neq b\\
    &\lbi=f(x), \ubi = f(x), && x=a \text{ or }  b
  \end{aligned} \right.
\label{eq:ublb}
\end{equation} 
By definition, regular points in $\phi^{-1}(e)$ share the same lower and upper bounds\myedit{, whereas critical points retain their original values}. 
See \cref{fig:ct-segmentation}(B) for an example, where regular points in the pink region use $f(b)$ and $f(a)$ as their initial lower and upper bounds.
For all datapoints $\{x_1, \dots, x_N\}$ in the domain, we define   their pointwise upper and lower bounds as high-dimensional vectors, {\ie}, $\ub = (\ubi_1, ..., \ubi_N)$ and $\lb = (\lbi_1, ..., \lbi_N)$.  

\para{A toy example.}
We use $f$ from \cref{fig:ct} as an example with a persistence threshold  $\vareps=0.1$.  
The initial upper and lower bounds $\ub$ and $\lb$ are computed via Eqn.~\eqref{eq:ublb} based on the simplified contour tree $T_\vareps$. 
The boundaries of $\ub$ and $\lb$ follow the boundaries of the topological regions induced by  $T_{\vareps}$ (\myedit{{\cf} \cref{fig:ct-segmentation}A, assuming $T$ to be $T_{\vareps}$ in~\cref{fig:ct-segmentation}C}). 
Using $\ub$ and $\lb$ as pointwise error bounds for TopoSZ, we study the decompressed scalar field $f'_{\vareps}$. 
We observe two FPs in the contour tree $T'_\vareps$ of $f'_{\vareps}$. 
Indeed, using the customized SZ compressor with initial pointwise lower and upper bounds does not necessarily eliminate all false cases.
Our next step is to iteratively apply the customized SZ by updating the upper and lower bounds until no false cases are detected.

\subsubsection{Eliminating False Cases With Iterations} 
\label{sec:iteration}

Before introducing our iterative strategy, we first define the multilayer neighborhoods around a given point $x \in \Xspace$, inspired by the multilayer prediction model in~\cite{TaoDiChen2017}. 

\para{Multilayer neighborhoods.} 
Consider a 2D scalar field defined on a uniform grid. The multilayer neighborhoods of a datapoint $x$ is defined such that the 0-layer neighborhood of $x$ is the datapoint itself, the 1-layer neighborhood includes its 8 neighbors, and so on. A datapoint could grow its $k$-layer neighborhood (for any positive integer $k$) until it reaches the domain boundary.  


\para{1st iteration: dealing with false positives.}
During the 1st iteration, if an FP is detected from the decompressed data $f'_{\vareps}$, we find its corresponding edge in $T'_\vareps$ that is not in $T_\vareps$. 
As shown in \cref{fig:iteration-FP}(D), an FP occurs at $e' \in T'_\vareps$ (\cf,~\cref{fig:ct}C). 
We then update the upper and lower bounds for a subset of datapoints in $\Xspace$ for the 1st iteration as follows:

\begin{figure}[t]
    \centering
    \includegraphics[width=1.01\columnwidth]{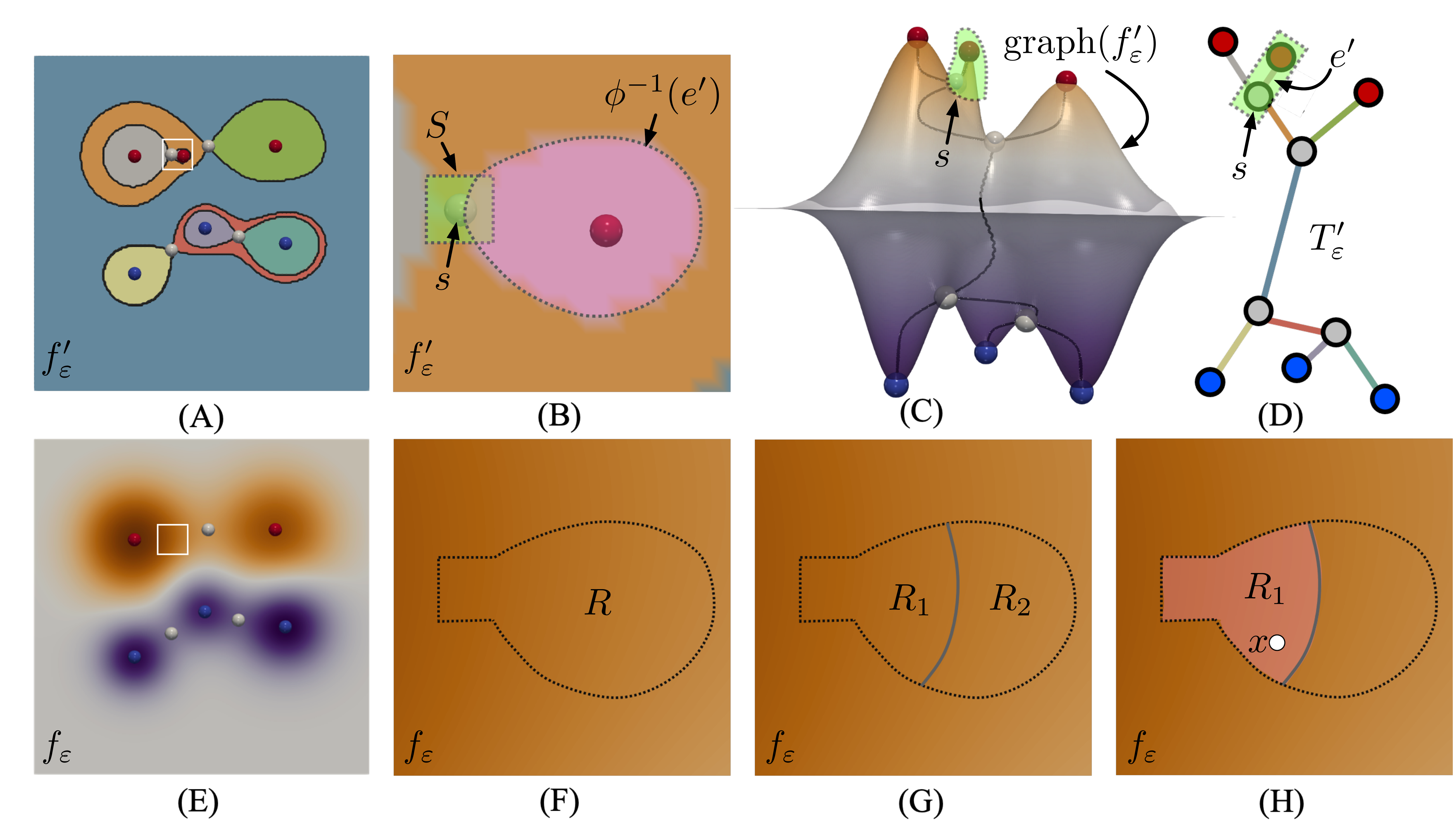}
    \vspace{-6mm}
    \caption{Updating the lower and upper bounds when an FP occurs. (A) A decompressed scalar field $f'_{\vareps}$ colored by topological regions induced by $T'_{\vareps}$. (B) The zoomed-in view of the white square from (A), where an FP occurs. (C) The graph of $f'_{\vareps}$. (D) The contour tree $T'_\vareps$. (E) The original simplified scalar field $f_{\vareps}$ colored by the $f_{\vareps}$ values. (F) The zoomed-in view of the white square from (E) that contains the region $S$ where the upper and lower  bounds need to be updated. (G) A refinement of $S$ into two monotonic regions according to P2. (H) The lower and upper bounds for $x$ are decided by the partition $x$ belongs to.} 
    \vspace{-2mm}
    \label{fig:iteration-FP}
\end{figure}

\para{\underline{P1.}}
Let $s$ be a point in $\Xspace$ that maps to the saddle point of $e'$. 
We first combine $s$ with its 1st-layer neighborhood to form a region $S$; see the green square of \cref{fig:iteration-FP}(B). 
We then combine $S$ with the pre-image $\phi^{-1}(e')$ (pink region in \cref{fig:iteration-FP}B) to form a region $R$; see \cref{fig:iteration-FP}(F). 
For all datapoints in $R$, we need to update their upper and lower bounds using their values in the original simplified scalar field  $f_\varepsilon$ to eliminate the FP involving $s$.   
Notice that $s$ is a saddle point in $f'_{\vareps}$ but a regular point in $f_{\vareps}$. 

\para{\underline{P2.}}	
Since this is the 1st iteration, we refine the region $R$ by evenly dividing the datapoints in $R$ into two subregions $R_1$ and $R_2$ based on the original scalar field $f_{\vareps}$.
Let $\min(R)$ and $\max(R)$ denote the minimum and maximum  $f_{\vareps}$ values among datapoints in $R$. 
The refinement of $R$ satisfies two conditions: (1) $R_1$ and $R_2$ are monotonic in $f$; and (2) $\max(R_1) \leq \min(R_2)$; see \cref{fig:iteration-FP}(G). 

\para{\underline{P3.}}
For any regular point $x$ in the refined region $R_1$, its new upper bound and lower bound are updated to be $u := \max(R_1)$ and $l := \min(R_1) = f(s)$; see~\cref{fig:iteration-FP}(H). The upper and lower bounds for datapoints in $R_2$ are updated similarly.  

\begin{figure}[!ht]
    \centering
    \includegraphics[width=1.0\columnwidth]{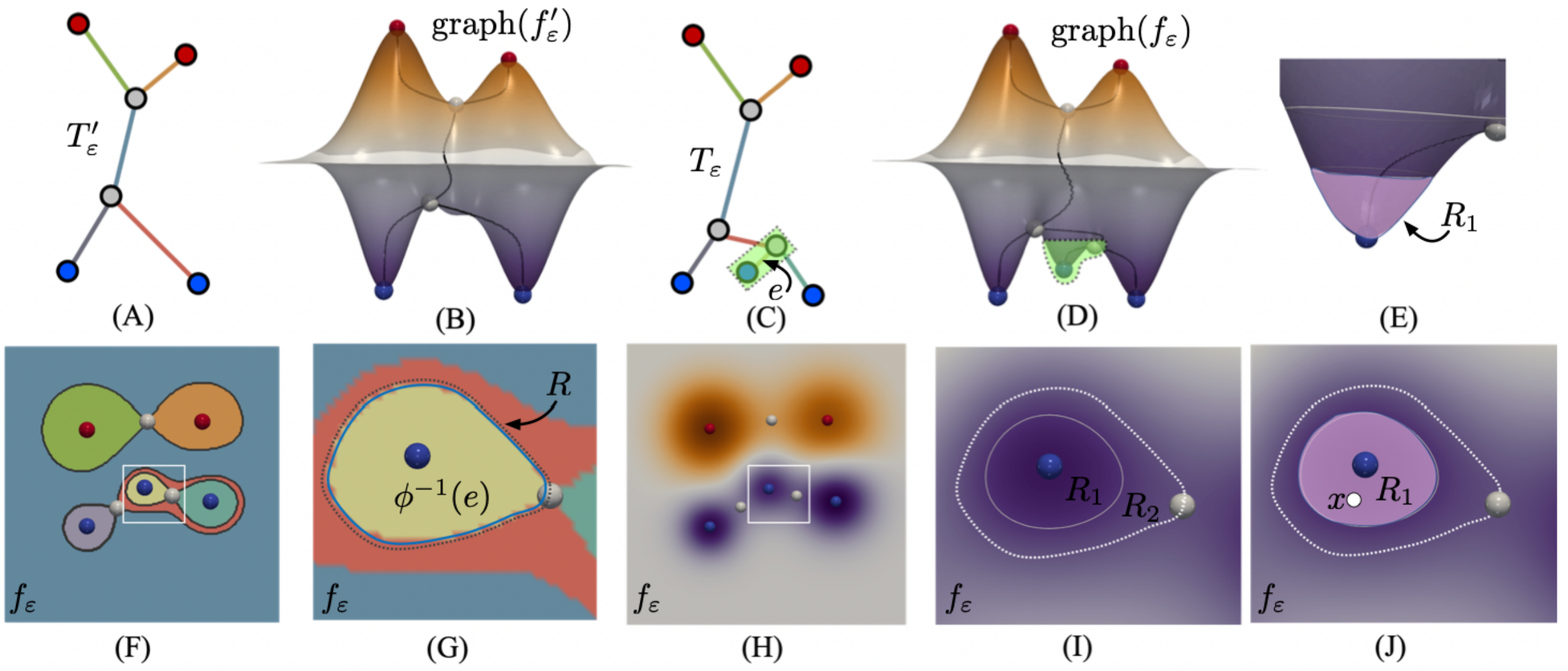}
    \vspace{-6mm}
    \caption{Updating the lower and upper bounds when an FN occurs. The contour tree $T'_{\vareps}$ (A) and the graph (B) of a 2D decompressed scalar field $f'_{\vareps}$. The contour tree $T_{\vareps}$ (C) and the graph of a 2D scalar field $f_\vareps$ (D), together with the $T_{\vareps}$ induced  segmentation (F). (G) The zoomed-in view of (F), where an FN occurs. (I) A refinement of the region $R$ according to P2. (E), (I) and (J) are the zoomed-in views of the white box in (H). The  lower and upper bounds for $x$ are decided by the partition $x$ belongs to.} 
    \label{fig:iteration-FN}
\end{figure}

\para{1st iteration: dealing with false negatives.}
During the 1st iteration, if an FN is detected from $f'_{\vareps}$, we find the corresponding edge $e \in T_\vareps$ that is missing in $T'_\vareps$. 
As shown in \cref{fig:iteration-FN}(A), an FN occurs \wrt~\cref{fig:iteration-FN}(C).
We then update the upper and lower bounds for a subset of datapoints in $\Xspace$ as follows:

\para{\underline{N1.}} 
Let $R$ represents the set of datapoints that combines the pre-image $\phi^{-1}(e)$ (yellow region) and their 1-layer neighborhood; see the region enclosed by a dotted line in \cref{fig:iteration-FN}(G).

\para{\underline{N2.}} 
We refine $R$ into two regions $R_1$ and $R_2$ following P2; see \cref{fig:iteration-FN} (I) for an example. 
We then update the upper and lower bounds for datapoints in $R$ following P3. For example, any regular point $x$ in the refined region $R_1$ has updated upper and lower bounds as  $u := \max(R_1)$ and $l := \min(R_1)$;  see \cref{fig:iteration-FN}(E) and (J). 

\para{1st iteration: dealing with false types.} 
During the 1st iteration, if an FT is detected from $f'_{\vareps}$, we find the corresponding edge $e \in T_\vareps$ whose end point (a local extremum) has a wrong type.  
An FT is dealt with similarly to the scenario involving a FN and is therefore omitted from the discussion. 

\para{k-th iteration: dealing with false cases.}  
We have described the region refinement process for the 1st iteration. 
During the $k$-th iteration (for $k \geq 1$), to eliminate false cases, we will consider the $k$-layer neighborhood of $s$ in P1 as well as the $k$-layer neighborhood of $\phi^{-1}(e)$ in N1. 
We then refine the region $R$ in P2 by dividing it into $k+1$ subregions $R_1, \dots, R_{k+1}$ such that (1) each $R_i$ is monotonic in $f$, for $1 \leq i \leq k+1$; and (2) $\max(R_i) \leq \min(R_{i+1})$, for $1 \leq i \leq k$. 
We update the upper and lower bounds in P3 according to these $k+1$ subregions. 
By design, the refinement process gets aggressive over time as $k$ increases to eliminate false cases. 
\myedit{We provide a discussion on the worst-case performance of iterations in~\cref{sec:worst-case}. }

\subsection{A Customized Error-Controlled Quantization}
\label{sec:quantization}
To incorporate topological constraints---defined by the pointwise upper bound $\ub:=[u_1, \dots, u_N]$ and the lower bound $\lb:=[l_1, \dots, l_N]$---into the data compression process, we propose a novel error-controlled quantization encoder for TopoSZ by modifying the one from the SZ-1.4 in ~\cref{sec:classicSZ} (see a detailed description in~\cite{TaoDiChen2017}).

For {\tool}, assume we are given the global error bound $\xi$ and a persistence threshold $\vareps$. 
For a datapoint $x \in \Xspace$ with an original value $f(x)$, we first   calculate its lower error bound $l$ and upper error bound $u$ using the approach described in \cref{sec:ublb}. 

\begin{figure}[!ht]
    \centering
    \vspace{-2mm}
    \includegraphics[width=\columnwidth]{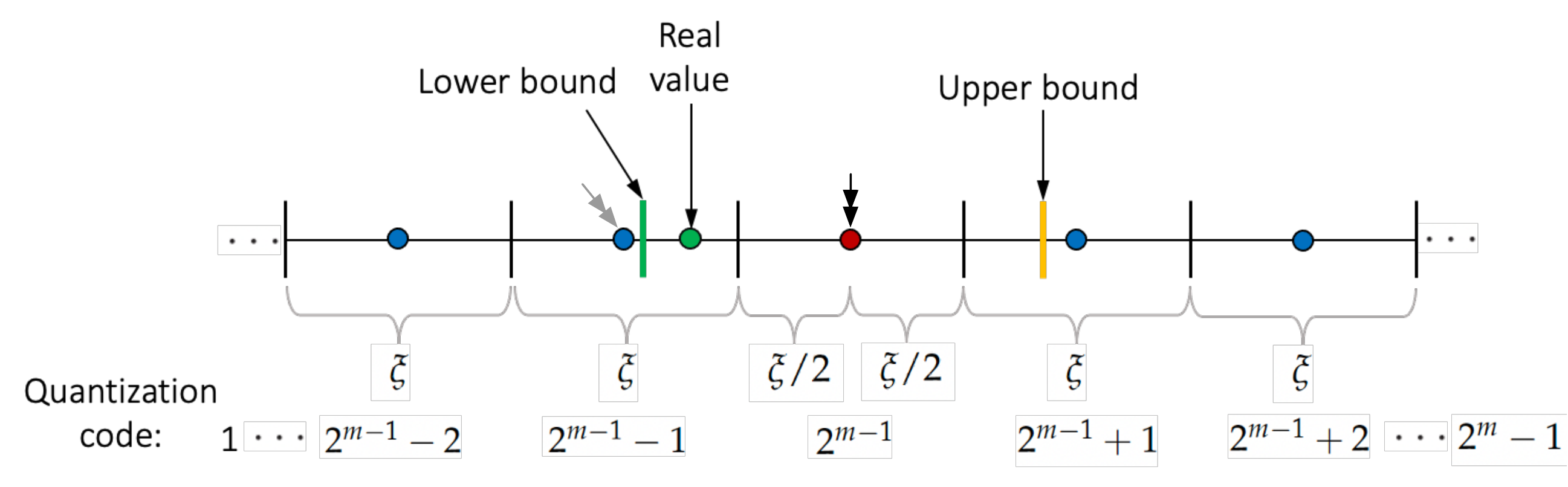}
    \vspace{-8mm}
    \caption{A customized error-controlled quantization encoder for {\tool} based on the upper and lower bounds.} 
    \label{fig:error-control}
     \vspace{-2mm}
\end{figure}

In the original work of Tao {\etal}~\cite{TaoDiChen2017}, the distance between any two adjacent 2nd-phase predicted values equals $2\xi$; however, in our setting, this distance is set to be $\xi$. 
In {\tool}, we need to consider the location of $l$ (green bar) and $u$ (yellow bar) {\wrt} to the $2^m-1$ intervals that contain 1st- and 2nd-phase predicted values. Any predicted value smaller than $l$ or greater than $u$ is not considered. 
As illustrated in~\cref{fig:error-control}, when the real value of the datapoint $f(x)$ (green point) falls into a certain interval, the predicted values that are close to the real value in both directions are considered (gray and black double arrows). 
In the illustrated example, both predicted values in the $(2^{m-1}-1)$-th and $2^{m-1}$-th intervals serve as candidates, and both are less than $\xi$ away from the real value. 
However, only the predicted value in the $(2^{m-1})$-th interval falls inside $[l,u]$ (black double arrow) and is selected.  
Therefore, $x$ will have $2^{m-1}$ as its quantification code under TopoSZ. 
An unpredictable datapoint and its quantification code are sent to a lossless compressor for further compression.

\begin{table*}[!ht]
\caption{Comparing run time between {\tool} and other lossy-compressors, including SZ3, ZFP, FPZIP, TTHRESH, and TopoQZ. Compression of {\tool} contains two parts: initialization and iteration. CT, UBLB, and CSZ-1.4 represent the run time for computing the simplified contour tree \myedit{with its induced segmentation}, updating the lower and upper bound, and applying a customized SZ-1.4. \myedit{$\#$ means the number of iterations}. CM and DC stand for compression and decompression, respectively. All experiments use a global error bound $\xi = 0.06$. For {\tool} and TopoQZ, the persistence threshold $\vareps=0.06$. All times are in seconds.} 
\label{table:all-time}
\centering
\resizebox{\textwidth}{!}{
\begin{tabular}{lllllllll|llllllllll}
\toprule
   \multirow{2}{*}{\textbf{Dataset}} & \multicolumn{3}{c}{\textbf{Initialization}} & \multicolumn{4}{c}{\textbf{Iteration}} &
   \multirow{2}{*}{\textbf{DC}}
   & \multicolumn{2}{c}{\textbf{SZ3}}
   & \multicolumn{2}{c}{\textbf{ZFP}} 
   & \multicolumn{2}{c}{\textbf{FPZIP}}
   & \multicolumn{2}{c}{\textbf{TTHRESH}}
   & \multicolumn{2}{c}{\textbf{TopoQZ}}\\
  \cmidrule(lr){2-4}\cmidrule(lr){5-8}\cmidrule(lr){10-11}\cmidrule(lr){12-13}\cmidrule(lr){14-15}\cmidrule(lr){16-17}\cmidrule(lr){18-19}
   &\textbf{CT}&\textbf{UBLB} & \textbf{CSZ-1.4} &\textbf{CT}&\textbf{UBLB} & \textbf{CSZ-1.4}&\textbf{\#}&&\textbf{CM}&\textbf{DC}&\textbf{CM}&\textbf{DC}&\textbf{CM}&\textbf{DC}&\textbf{CM}&\textbf{DC}&\textbf{CM}&\textbf{DC}\\
\midrule
	$\HF$ & 3.59 & 0.06 & 0.01 & - & - & - & 0 & 0.003&0.003&7.9e-4&0.001&9.1e-4&0.002&2.09e-3&0.20&0.005&3.12&1.68\\
	$\EW$ & 12.18 & 0.93 & 0.04 & 7.09 & 0.50 & 0.04 & 4&0.02 &0.03&0.01&0.02&0.02&0.04&0.04&2.73&0.36&5.11&3.93\\
	$\TN$ &  51.13 & 1.78 & 0.07 & - & - & - & 0 &0.03&0.05&0.02&0.03&0.02&0.05&0.06&0.65&0.47&7.13&4.57\\
	$\VF$ & 21.68 & 1.83 & 0.04 & 11.76 & 0.95 & 0.07 & 7 &0.04&0.06&0.02&0.03&0.02&0.06&0.07&0.67 &0.51&6.70&4.29\\
	$\AF$ & 83.79 & 5.62 & 0.20 & 49.53 & 2.61 & 0.20 & 3 &0.1& 0.19&0.08& 0.11&0.06& 0.13&0.14 &1.72&0.96&27.36&12.22\\
    $\IB$ & 286.8 & 19.31 & 0.74 & 142.7 & 8.78 & 0.72 & 1 &0.31&0.51&0.18& 0.32&0.14&0.54&0.62& 9.08&4.80&114.95&44.39\\
\bottomrule
\end{tabular}
}
\end{table*}

\section{Experimental Results}
\label{sec:results}

\begin{figure}[!ht]
 \centering
  \includegraphics[width=1.0\linewidth]{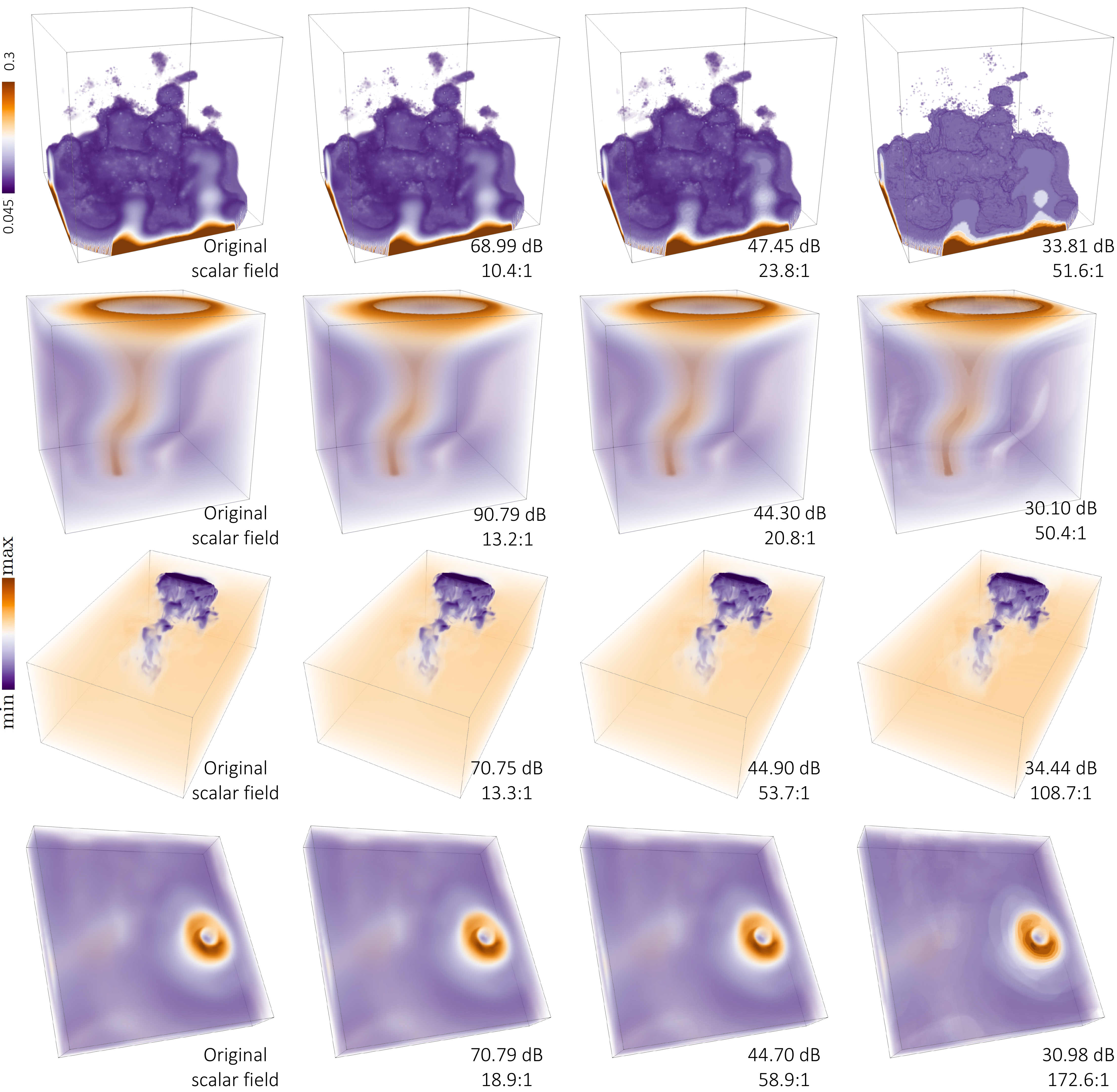}
  \vspace{-2mm}
  \caption{
    3D visualization for the $\VF$ (1st row), $\TN$ (2nd row), $\AF$ (3rd row), and $\IB$ (4th row) datasets. Columns 1-4: the original data, the decompressed data with high, medium, and low quality, respectively. We also report the PSNR and the compression ratio for each decompressed dataset.
  }
  \label{fig:datasets-3D}
\end{figure}

We present experimental results using two 2D and four 3D scientific datasets described in Table~\ref{table:data}. 
The details of these datasets are described in \cref{sec:data-details}. 

\begin{table}[!ht]
\caption{Datasets used in our experiments.} 
\label{table:data}
\vspace{-2mm}
\centering
\begin{tabular}{|l|l|l|}
\hline
\textbf{Dataset} &\textbf{Dimensions }&\textbf{Size}\\
\hline
	$\HF$ & $150 \times 450 $ & 270 KB\\
	$\EW$ & $1440\times 720$& 4.1 MB \\
	$\VF$ & $128 \times 128 \times 128$ & 8.4 MB\\ 
	$\TN$ & $128\times 128 \times 128$ & 8.4 MB\\ 
	$\AF$ & $ 300\times 180 \times 120$ & 25.9 MB\\
    $\IB$ & $500\times 500 \times 90$ & 90 MB\\
\hline
\end{tabular}
\end{table}

\para{Preprocessing.}
For each dataset, the original scalar field is normalized to $[0,1]$ and saved using 32-bit floating points.   
Such a normalization helps us specify the persistence simplification level and pointwise error bound. 
For example, $\vareps=0.1$ means a $10\%$ persistence simplification of the range of the scalar field, and $\xi=0.01$ means a $1\%$ maximum pointwise error bound. 
For computing the contour tree and applying persistence simplification, we use algorithms by Tierny and Pascucci~\cite{TiernyPascucci2012} and Gueunet~\etal~\cite{GueunetFortinJomier2019}, with  implementations available in the Topology ToolKit (TTK)~\cite{TiernyFavelierLevine2017}. 
All results are obtained on a desktop computer with an Intel Core i7 CPU (2.8 GHz, 4 cores) and 16 GB RAM. 

\para{Evaluation metrics.} We explore a number of  metrics, including the number of false cases (in the decompressed data), data compression ratio, PSNR, bottleneck~\cite{Cohen-SteinerEdelsbrunnerHarer2007} and Wasserstein distances~\cite[page 183]{EdelsbrunnerHarer2010} between $f$ and $f'$; see the supplementary material for a review. 

\para{Overview of results.}
We first give a snapshot of compression capabilities of {\tool} in~\cref{sec:result-EW}, in comparison with other error-bounded compressors such as \myedit{SZ-1.4}, SZ3, ZFP, FPZIP, and TTHRESH; since these compressors are topology-agnostic, they typically lead to worse topology preservation by nature. 
\cref{fig:false-cases} shows that {\tool} consistently preserves topology of the input scalar field without introducing false cases. 
We then dive deeper into the algorithmic process of {\tool} by visualizing intermediate decompressed data during iterations (\cref{sec:results-iterations}), and study how its performance is affected by the parameter setting  (\cref{sec:performance}). 
We then compare {\tool} against TopoQZ, the compression framework most relevant to ours that  also offers some topological guarantees (see  \cref{sec:compare}). 
We conclude with a detailed run time analysis of {\tool} in \cref{sec:runtime}.
\myedit{See~\cref{sec:nyx} for an analysis of the compression quality {\wrt} the size of data using larger datasets in comparison with those described in Table~\ref{table:data}.} 

\subsection{A Snapshot of Compression Capabilities}
\label{sec:result-EW}

\begin{figure}[!ht]
    \centering
    \includegraphics[width=\columnwidth]{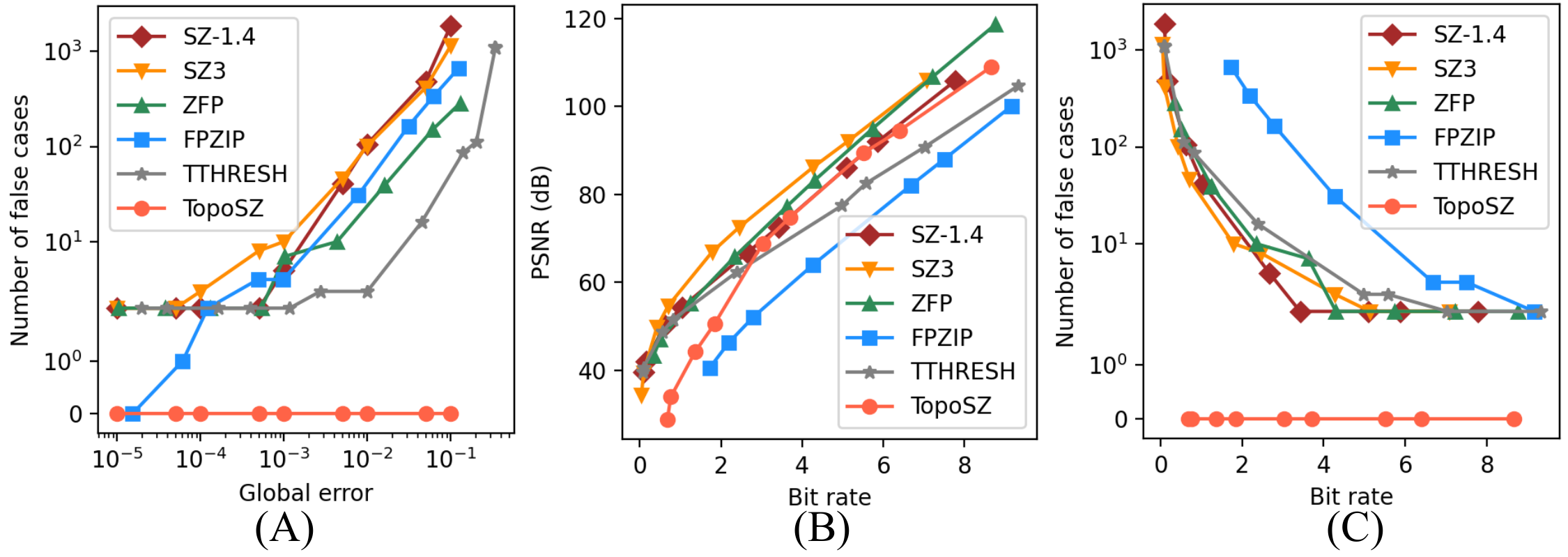}
    \vspace{-6mm}
    \caption{Test lossy compressors with the $\VF$ dataset: (A) the number of false cases \wrt~global error bound; (B) and (C): the PSNR and the number of false cases \wrt~bit rate (i.e.,~average bits per compressed data sample). Lossy compressors include {\tool}, \myedit{SZ-1.4}, SZ3, ZFP, FPZIP, and TTHRESH.} 
    \label{fig:false-cases}
    \vspace{-2mm}
\end{figure}

\cref{fig:false-cases} gives a snapshot of the compression results on the \myedit{$\VF$} dataset, comparing {\tool} against other state-of-the-art lossy compressors that do not focus on topology preservation  (i.e.,~\myedit{SZ-1.4}, SZ3, ZFP, FPZIP, and TTHRESH). 
We set a persistence threshold $\vareps=0.12$. 
\myedit{For experimental purposes, we can use any persistence threshold. However, we recommend setting $ \xi > \vareps$, since a smaller global error bound leads to smaller topological regions that need fine-grained control, and, therefore, less iterations and a higher compression ratio. We choose the current threshold that preserves a sufficient number of critical points while eliminating a certain amount of noise in the data, and is slightly larger than the largest global error bound tested in this experiment; see~\cref{sec:sensitivity} for a detail discussion on the persistence threshold and compression capabilities of {\tool}.} 
As shown in~\cref{fig:false-cases} (A), {\tool} generates a decompressed scalar field that preserves the simplified contour tree from the original scalar field without any false cases.  
In contrast, there are false cases within the decompressed data using other compressors, unless we specify a very small global error bound (e.g.,~$10^{-5}$ for FPZIP). 
These small error bounds indicate lossless compression for these off-the-shelf compressors. 

We also present the rate-distortion plots of all the evaluated compressors in~\cref{fig:false-cases} (B) and (C), using PSNR and the number of false cases as the distortion metrics, respectively. 
Rate-distortion is a widely used assessment metric in the compression community, where rate represents the average number of bits in the compressed data (e.g., it can be computed by 32 over the compression ratio for single-precision floating point data). 
According to these two figures, TopoSZ has a slightly worse rate distortion in terms of PSNR compared to SZ3 and ZFP, but it outperforms all the error-bounded compressors on preserving topology.

\subsection{Improved Compression Quality with Iterations}
\label{sec:results-iterations}

{\tool} couples a pointwise error-controlled compressor with topological constraints. 
It repeatedly applies a customized SZ (see~\cref{sec:quantization}) with updated lower and upper error bounds until no false cases are detected. 
We demonstrate that {\tool} improves the compression quality during iterations using the $\HF$, $\EW$, \myedit{and $\VF$} datasets.

\begin{figure}[t]
    \centering
    \includegraphics[width=\columnwidth]{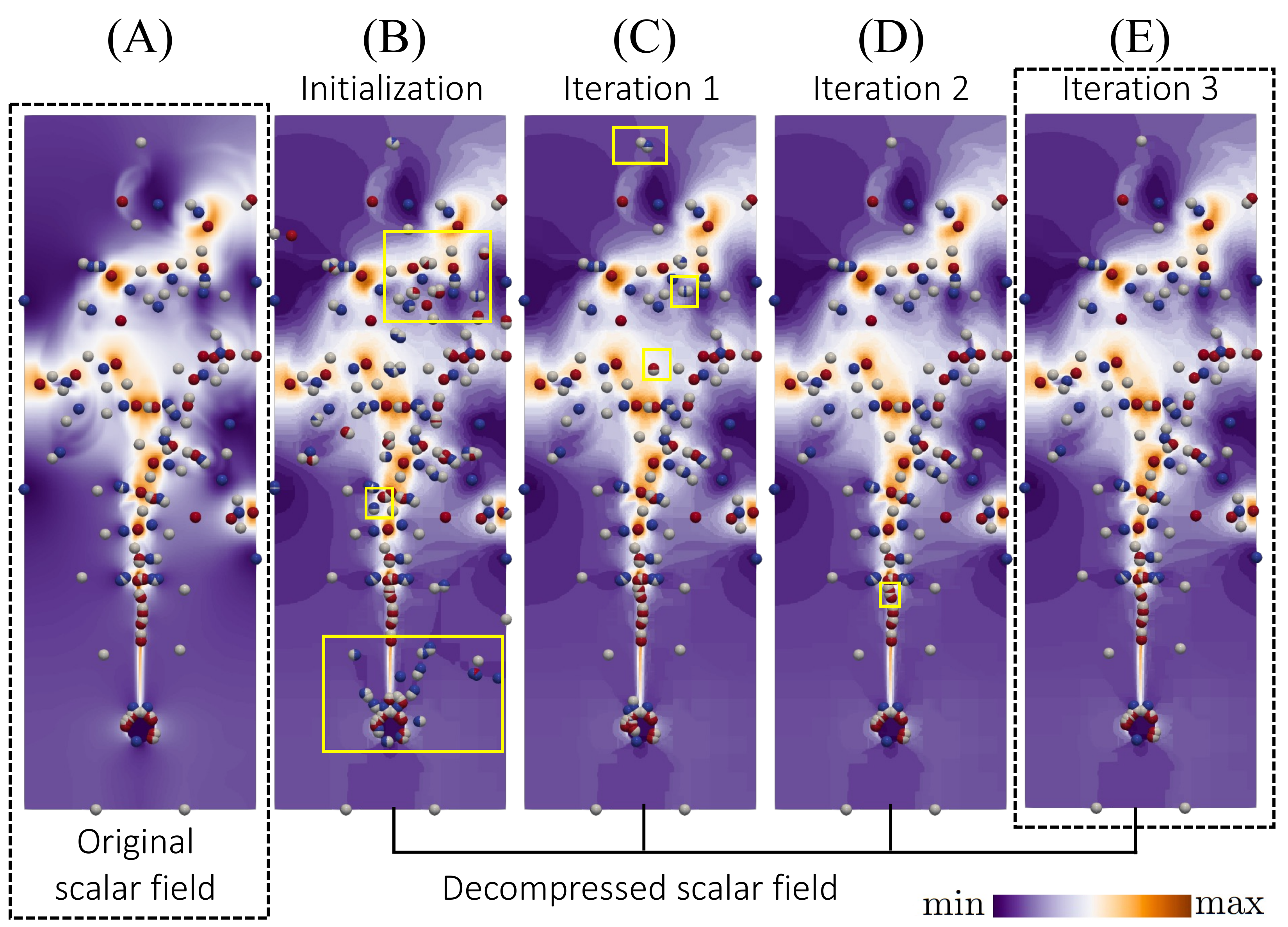}
    \vspace{-2mm}
    \caption{\myedit{The compression process of the $\HF$ dataset, with a persistence threshold $\vareps=0.02$ and a global error bound $\xi=0.05$. (A)-(E): the original scalar field, the decompressed scalar fields in the initialization step, 1st iteration, 2nd  iteration, and 3rd iteration, respectively. Local maxima are in red, local minima are in blue, and saddles are in white. Yellow boxes contain  false cases.}} 
    \label{fig:HF}
\end{figure}

First, {\tool} eliminates false cases during iterations. 
As shown in~\cref{fig:HF}, given a persistence threshold $\vareps=0.02$ and global error bound $\xi=0.05$, the initialization gives rise to a number of false cases with the initial upper and lower bounds (F); some of these false cases are enclosed in the yellow boxes in (B). 
With updated upper and lower bounds (see \cref{sec:iteration}), the 1st iteration eliminates most false cases; {\cf} (B) and (C).  
However, a few false positives remain in the yellow box of (C) after the 1st iteration.  
Therefore, {\tool} goes through two more rounds of iterations with updated error bounds. 
For the {\HF} dataset, {\tool} eliminates all false cases after iteration 3, thus preserving the types and positions of local extrema from the original data (E). 

\begin{figure*}[t]
    \centering
    \includegraphics[width=2.0\columnwidth]{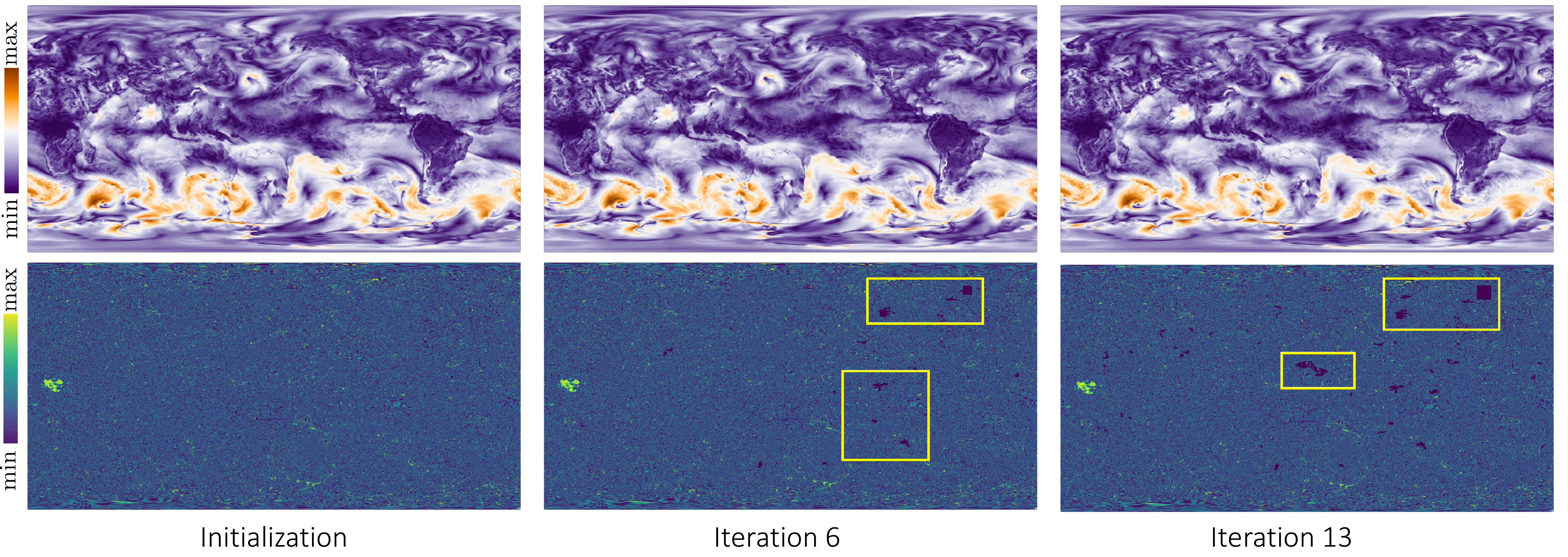}
    \vspace{-4mm}
    \caption{The compression process of the $\EW$ dataset, with a persistence threshold $\vareps=0.06$ and a global error bound $\xi=0.02$. Top: decompressed scalar fields during the initialization, the 6th iteration, and the 13th iteration, respectively. Bottom: pointwise error during the corresponding iterations.} 
    \label{fig:iteration-EW}
    \vspace{-2mm}
\end{figure*}

Second, {\tool} provides a finer grained control of pointwise errors between the original and the  decompressed data during each iteration. 
We apply {\tool} to the $\EW$ dataset with $\vareps=0.06$ and $\xi=0.02$.  
As shown in~\cref{fig:iteration-EW}, we visualize the decompressed data and its absolute pointwise differences {\wrt} the original data for selected iterations.
We observe that, with an increasing number of iterations, certain regions in the domain obtain lower pointwise errors compared with the previous iterations, for instance, the ones enclosed by the yellow boxes. 
{\tool} updates the lower and upper bounds wherever  false cases occur during iterations, and these updated bounds help decrease the errors within these regions in the decompressed data. 
Consequently, {\tool} improves the compression quality in terms of pointwise error during iterations.

\myedit{Third, we provide a detailed analysis of compression quality during iterations with the $\VF$ dataset ($\xi=0.006$ and $\vareps=0.06$) in Table~\ref{table:iterations}. The  number of false cases generally decreases with iterations, whereas PSNR remains unchanged. This is because we only need to update the lower and upper bounds within a small region during iterations ($<0.1\%$). These iterations effectively eliminate the detected false cases, but they cannot improve PSNR since the affected areas (by refinement) are too small. 
The compression ratio fluctuates with iterations, but it generally decreases. We notice a large decrease at the 8th  iteration using the $8$-layer neighborhood. Generally, a larger affected region leads to a larger decrease in the compression ratio and the number of false cases; see the 2nd and 8th iterations in Table~\ref{table:iterations}.}

\begin{table}[!ht]
\caption{\myedit{The analysis of compression quality during iterations for the $\VF$ dataset. Init. and Iter. represent the initialization and iteration steps. $\#$FC, EB ($\%$), and RO represent the number of false cases after each step, updated upper and lower bound {\wrt} the data domain ($\%$), and the compression ratio, respectively.}} 
\label{table:iterations}
\centering
\resizebox{\columnwidth}{!}{
\begin{tabular}{ccccc|ccccc}
\toprule
  &\textbf{\# FC} & \textbf{PSNR} &\textbf{EB ($\%$)} &\textbf{RO}&  &\textbf{\# FC} & \textbf{PSNR} &\textbf{EB ($\%$)}&\textbf{RO}\\
\midrule
	\textbf{Init.} & 15 & 53.9 & 100  & 18.0 & \textbf{5th} & 3 & 53.9 & 0.021 &17.9\\
	\textbf{1st} & 10 & 53.9 & 0.032 & 18.1 & \textbf{6th}& 1 & 53.9& 0.052 &17.8\\
	\textbf{2nd} &10 & 53.9 & 0.039 & 18.2 & \textbf{7th} & 2 & 53.9 & 0.035&17.9\\
	\textbf{3nd} &5 & 53.9 & 0.068 & 17.9 & \textbf{8th}& 0 & 53.9 & 0.070&17.5\\
	\textbf{4th} &4 & 53.9 & 0.005 & 18.1 & &  &  & &\\
\bottomrule
\end{tabular}
}
\end{table}

\subsection{Compression Performance}
\label{sec:performance}

{\tool} utilizes two parameters---a persistence threshold $\vareps$ and a global error bound $\xi$---to control the compression performance. 
We visualize four 3D datasets before and after compression with three levels of compression quality in~\cref{fig:datasets-3D}, that is, at high, medium, and low compression quality.  
We also display the PSNR and compression ratio next to each decompressed scalar field. 
As shown~\cref{fig:datasets-3D}, we obtain lower compression rates when preserving finer scale topological structures, whereas we obtain higher compression rates with low PSNR measurements. 
We further discuss how $\vareps$ and $\xi$ affect the performance of {\tool}.

\subsubsection{Performance Analysis {\wrt} Global Error Bound}
\label{sec:global-error}

\begin{figure*}[t]
    \centering
    \includegraphics[width=2\columnwidth]{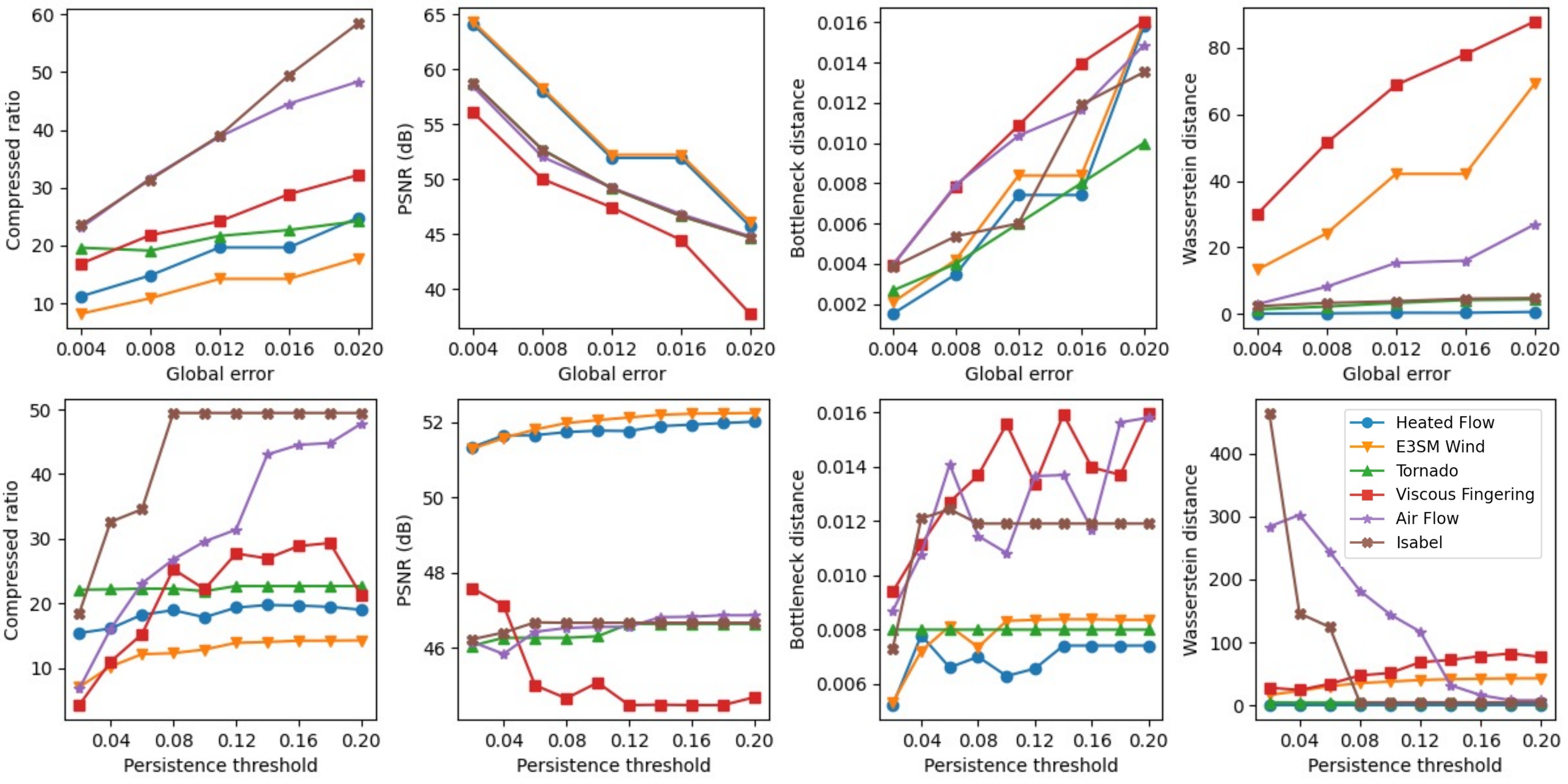}
    \vspace{-2mm}
    \caption{Performance analysis of {\tool} {\wrt} a global error bound $\xi$ and a persistence threshold $\vareps$. Top row: compression performance with an increasing $\xi$; $\vareps=0.12$. Bottom row: compression performance with an increasing $\vareps$; $\xi=0.012$. Columns 1-4: the compression ratio, PSNR, bottleneck distance, and Wasserstein distance, respectively. } 
    \label{fig:alldata}
    \vspace{-2mm}
\end{figure*}

With a fixed persistence threshold $\vareps=0.12$, we apply {\tool} to all datasets in Table~\ref{table:data} with a global error bound $\xi \in \{0.004, 0.008, 0.012, 0.016, 0.02\}$. 
The 1st row of~\cref{fig:alldata} shows the evolution of compression ratios, PSNRs, bottleneck distances, and Wasserstein distances as $\xi$ increases. 
All datasets exhibit similar trends in the compression rates, PSNRs, and bottleneck distances as $\xi$ increases. 
These trends imply that a higher $\xi$ value poses a more relaxed constraint and leads to a higher  compression rate and lower compression quality. 
The Wasserstein distances also exhibit an increasing trend across all datasets; in particular, for the $\EW$, $\VF$ and $\AF$ datasets. 
Comparing across datasets: the $\HF$ dataset contains only a few topological features due to its small size and coarse resolution; the $\TN$ and $\IB$ datasets contain one main topological feature; the $\EW$, $\VF$ and $\AF$ datasets contain many more features that are evaluated in the Wasserstein metric.

\subsubsection{Performance Analysis {\wrt} Persistence Threshold}
With a fixed global error bound $\xi=0.012$, we apply  {\tool} to all datasets with a persistence threshold $\vareps \in [0.02, 0.04, \dots, 0.2]$ (with a step size of  $0.02$). 
The evolution of the compression performance for an increasing $\vareps$ is shown in~\cref{fig:alldata} (the 2nd row). 
We observe that, regardless of the $\vareps$ value, the compression performance has only a few changes  for the $\HF$, $\TN$, and $\EW$ datasets.
However, $\VF$ and $\AF$ datasets contain a large number of critical points and are more difficult to handle with an error-controlled quantization (see~\cref{sec:quantization}). 
Therefore, a relaxed topological constraint (indicated by a higher $\vareps$) discards fine-error controls on regions where critical points have low persistence, and achieves a higher compression ratio. 
The $\IB$ dataset contains a main feature (describing the eye and its surroundings) with high persistence and numerous features with low persistence. 
Therefore, the compression ratio increases with $\vareps$ varying from $0.02$ to $0.08$ and remains the same after $\vareps=0.08$.
In general, the choice of $\vareps$ value has less impact on the compression performance than the choice of $\xi$. 

The two series of experiments in~\cref{fig:alldata} suggest that we can use a global error bound $\xi$ to control the compression quality, and the persistence threshold $\vareps$ to add additional topological constraints to the decompressed data. 

\subsection{Comparison with TopoQZ}
\label{sec:compare}

We now compare {\tool} with TopoQZ~\cite{SolerPlainchaultConche2018}.
TopoQZ is most relevant to our approach as it is designed to explicitly enforce a topological control. 
Its implementation is available via TTK~\cite{TiernyFavelierLevine2017}. 

\begin{figure*}[t]
    \centering
    \includegraphics[width=2.0\columnwidth]{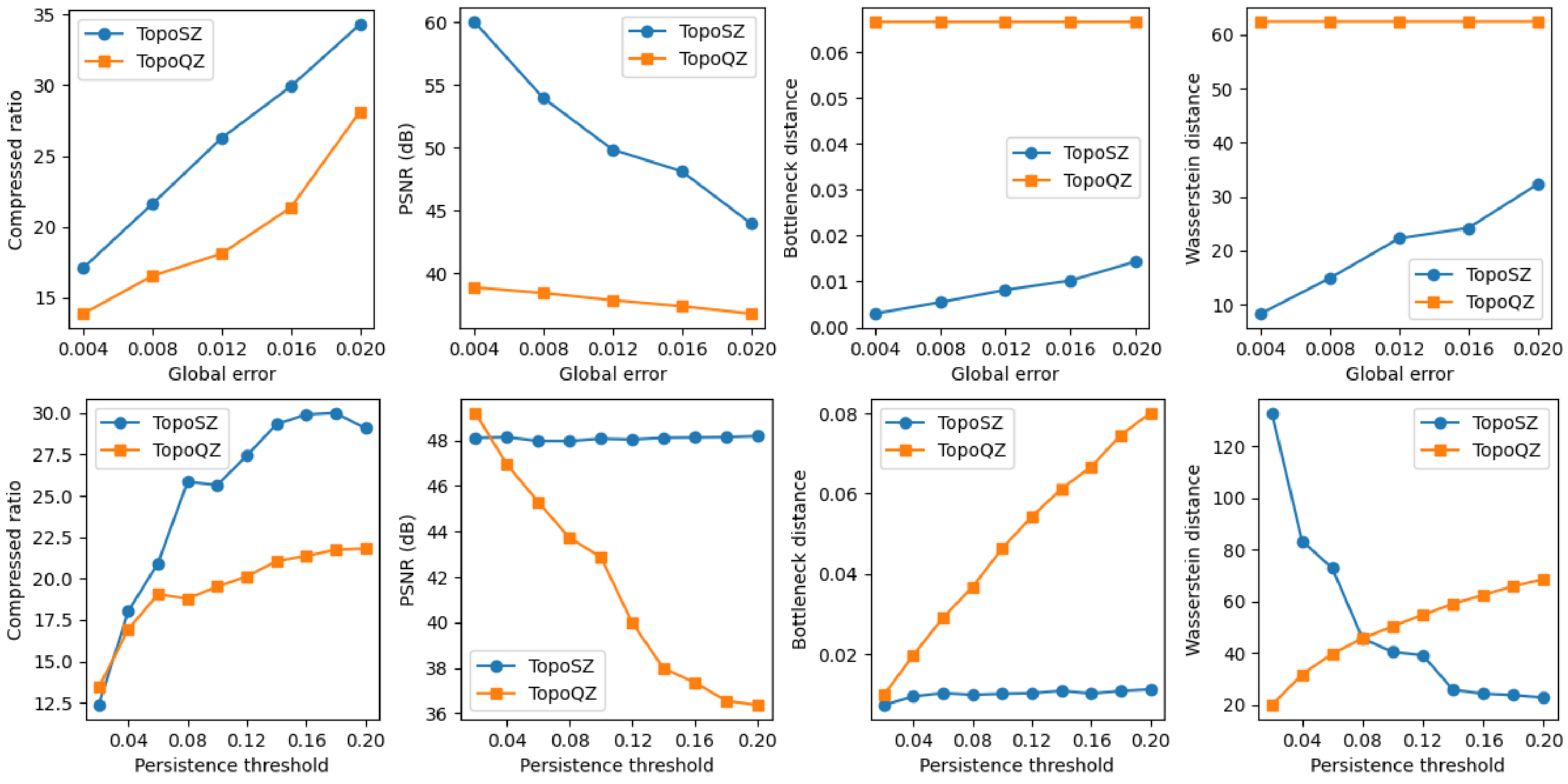}
    \vspace{-2mm}
    \caption{Comparing TopoSZ (blue) with TopoQZ (orange). Top row: compression performance with an increasing $\xi$; $\vareps=0.12$. Bottom row: compression performance with an increasing $\vareps$; $\xi=0.016$. Columns 1-4: the average  compression ratio, PSNR, bottleneck distance, and Wasserstein distance, respectively.} 
    \label{fig:compare-TopoQZ}
\end{figure*}

We rerun the two series of experiments from~\cref{fig:alldata} with TopoQZ.
We use the same parameter setting as {\tool}, that is, using a global error bound $\xi \in [0.004, 0.02]$ (with a step size of $0.004$) and a persistence threshold $\vareps \in [0.02, 0.2]$ (with a step size of $0.02$).
Using TopoQZ, we also obtain decompressed data without any false cases. 
We compare the performance of {\tool} with TopoQZ in~\cref{fig:compare-TopoQZ}.

\subsubsection{Comparing Pointwise Error Control}
Soler~\etal~claimed that TopoQZ can be extended to enforce pointwise error control~\cite{SolerPlainchaultConche2018}. 
However, their maximum pointwise error between the original and decompressed data is dependent on the persistence threshold. 
\myedit{\cref{sec:error-control} demonstrates the evolution of the pointwise error control between TopoQZ and {\tool}.}
It shows that, to the best of our knowledge, {\tool} is the first lossy compressor that combines pointwise error control and topological guarantee during compression. 

\subsubsection{Comparing Compression Performance} 

We provide the evolution of the compression ratios, PSNRs, bottleneck distances, and Wasserstein distances, again averaged over all datasets, for a varying $\xi$ or $\vareps$ using {\tool} (blue curves) and TopoQZ (orange curves) in~\cref{fig:compare-TopoQZ}.  These average curves are meaningful since all our datasets are normalized within $[0,1]$ during preprocessing. 

The 1st row of~\cref{fig:compare-TopoQZ} shows the compression performance with an increasing $\xi$ and a fixed $\vareps$. 
The row indicates that {\tool} has a better compression performance than TopoQZ in this experiment in terms of the compression ratio, PSNR, and topological preservation. 
Taking a closer look at PSNR, TopoQZ has a lower rate than {\tool}, and the bottleneck and Wasserstein distances do not change with a varying $\xi$ using  TopoQZ. 
Since TopoQZ does not have a strict control on the  pointwise error, an increasing $\xi$ has less impact on the compression performance of TopoQZ. 

On the other hand, as shown in the 2nd row of~\cref{fig:compare-TopoQZ}, with a fixed $\xi$, the compression performance of TopoQZ is highly dependent on the $\vareps$ values. 
However, {\tool} still exhibits better compression performance than TopoQZ under this setting for two reasons. First, data predictions from the SZ lead to the decompressed values being closer to the original values. 
Such a strategy also enforces a strict limitation on the decompressed data in the amount of deviation during compression. 
TopoQZ conducts a linear interpolation of the quantized simplified function, with a worse pointwise error when the dataset is nonlinear or irregular. 
Second, {\tool} yields a higher compression capability; see~\cref{fig:false-cases}. 

However, {\tool} appears to have a higher Wasserstein distance when $\vareps$ is small (\cref{fig:compare-TopoQZ} bottom right) because TopoQZ simply removes all contour tree branches associated with the original scalar field when their persistences are smaller than $\vareps$. Therefore, the Wasserstein distance for TopoQZ increases when $\vareps$ increases. 
{\tool}, on the other hand, predicts values based on their neighborhoods. With a small $\vareps$, the predictor produces numerous contour tree branches with small persistence.
Although they do not affect the PSNR or the bottleneck distance, they do affect the Wasserstein distance cumulatively. 

In conclusion, {\tool}, which is derived from the SZ-1.4, achieves a higher compression ratio than TopoQZ in these experiments.

\subsection{Run Time Analysis} 
\label{sec:runtime}

Table~\ref{table:detailed-runtime} provides a detailed run time analysis for {\tool}, which contains two key components, initialization (see \cref{sec:initialization}) and iteration (see \cref{sec:iteration}). 
\myedit{Average run time across iterations is reported for the iteration component.}
During initialization, we need to compute a simplified contour tree $T_\vareps$ \myedit{with its induced segmentation}, update the pointwise upper and lower bounds $\ub$ and $\lb$, and apply a customized SZ. 
We also need to check for false cases within the simplified contour tree $T_\vareps'$ from the decompressed data.
If there are false cases, we proceed with iterations, where we update the error bounds for a small subset of points. We need to check for false cases in $T_\vareps'$ at the end of each iteration.  
As shown in Table~\ref{table:detailed-runtime}, the average run time for each iteration is much less than the initialization since only a small set of points is affected to eliminate false cases. 
The run time bottleneck is computing the contour trees.  

\begin{table}[!ht]
\caption{A detailed run time analysis of {\tool} with a varying persistence threshold $\vareps$. \myedit{Notations in this table follow those of Table~\ref{table:all-time}. All times are in seconds.}} 
\label{table:detailed-runtime}
\centering
\resizebox{\columnwidth}{!}{
\begin{tabular}{cccccccccc}
\toprule
\multirow{2}{*}{\textbf{Dataset}} 
&\multirow{2}{*}{{$\boldsymbol{\vareps}$}}
&\multicolumn{3}{c}{\textbf{Initialization}} 
&\multicolumn{4}{c}{\textbf{Iteration}}\\
\cmidrule(lr){3-5}\cmidrule(lr){6-9}
&&\textbf{CT}&\textbf{UBLB} & \textbf{CSZ-1.4} 
&\textbf{CT}&\textbf{UBLB} & \textbf{CSZ-1.4}
&\textbf{\# Iterations}\\
\midrule
\multirow{3}{*}{$\HF$} & 0.01 & 3.85 & 0.07 & 0.03 & 2.03 & 0.03 & 0.03 & 1\\
	& 0.06 & 3.95 & 0.06 & 0.03 & - & - & - & 0\\
	& 0.12& 3.92 & 0.06 & 0.03 & - & - & - & 0\\
	\cmidrule(lr){1-9}
	\multirow{3}{*}{$\EW$} &0.01& 13.60 & 1.07 & 0.16 & 7.59 & 1.83 & 0.14 & 5\\
	&0.06 & 12.86 & 0.96 & 0.15 & 7.06 & 0.64 & 0.14 & 4\\
	&0.12 & 11.99 & 0.94 & 0.16 & 6.25 & 0.46 & 0.14 & 1\\
	\cmidrule(lr){1-9}
	\multirow{3}{*}{$\TN$} &0.01 & 49.49 & 2.00 & 0.31 & 14.20 & 2.16 & 0.22 & 4\\
	&0.06 & 51.13 & 1.78 & 0.26 & - & - & - & 0\\
	&0.12 & 24.21 & 1.85 & 0.26 & - & - & - & 0\\
	\cmidrule(lr){1-9}
	\multirow{3}{*}{$\VF$} &0.01 & 20.99 & 1.93 & 0.20 & 11.72 & 6.56 & 0.20 & 15\\
	&0.06& 21.01 & 1.85 & 0.24 & 11.67 & 1.50 & 0.21 & 8\\
	&0.12 & 21.43 & 1.84 & 0.23 & 11.45 & 0.87 & 0.20 & 5\\
	\cmidrule(lr){1-9}
	\multirow{3}{*}{$\AF$} &0.01 & 90.50 & 5.91 & 0.82 & 55.34 & 3.12 & 0.72 & 9 \\
	&0.06 & 83.79 & 5.62 & 0.72 & 49.53 & 4.11 & 0.62 & 3\\
	&0.12 & 79.09 & 5.55 & 0.71 & 45.00 & 6.37 & 0.64 & 2\\
	\cmidrule(lr){1-9}
    \multirow{3}{*}{$\IB$} &0.01 & 472.10 & 19.66 & 2.67 & 228.33 & 117.82 & 2.24 & 16\\
    &0.06 & 281.26 & 18.92 & 2.71 & 140.15 & 8.78 & 2.72 & 1\\
    &0.12& 263.29 & 19.09 & 2.59 & - & - & - & 0\\
\bottomrule
  \end{tabular}}
\end{table}

In addition, we investigate the run time of {\tool} with a varying persistence threshold $\vareps$. 
The global error bound $\xi$ has an impact on the iteration time, but it does not affect the run time for computing simplified contour trees. 
Let $\xi=0.06$ for all experiments in Table~\ref{table:detailed-runtime}. 
We vary the persistence threshold $\vareps=0.01, 0.06$, and $0.12$ for each dataset. 
Table~\ref{table:detailed-runtime} shows that the number of iteration decreases as we increase the persistence 	threshold. TTK also needs less time to compute the sampled contour trees with a large $\vareps$.

For comparison, we also provide run time analysis of SZ3, ZFP, FPZIP, TTHRESH, and TopoQZ, as shown in Table~\ref{table:all-time}.
All experiments are run with a global error bound $\xi=0.06$. 
For {\tool} and TopoQZ, we use a persistence threshold $\vareps=0.06$.
Notice that for these two topology preserving compressors, the majority (at least 90$\%$) of the compression time is spent on computing the contour trees. 
We argue that in certain use cases (e.g., mitigating the requirement on storage capacity), achieving a higher compression ratio is more important than compression speed. 
Also, scientific data are usually compressed once when they are generated and written to the storage systems, but may be decompressed multiple times when they are retrieved for various post hoc data analytics. TopoSZ delivers decent compression ratios and fast decompression speed while preserving important topological information, which helps mitigate the extreme-scale data challenges in many uses cases. 
Finally, our contour tree computation uses the implementation from TTK and has not been optimized for performance, which is left for future work.

\section{Conclusion}
\label{sec:conclusion}

We introduce {\tool}, a novel lossy compression framework that preserves topological features while enforces pointwise error control. 
In particular, {\tool} imposes upper and lower error bounds during compression using topological  information from contour trees. 
It achieves a pointwise error control based on a customized quantization derived from the SZ compressor (version 1.4). 
Our experimental results demonstrate the strengths of {\tool} in preserving the topological features while achieving good compression quality, in comparison to state-of-the-art lossy compressors with comparable compression rates. 
In general, any prediction-based error-bounded compressor may be customized using a similar strategy for topology-preserving compression, and it is feasible to extend {\tool} using a multilayer predictor~\cite{TaoDiChen2017}. 

For future work, we would like to further improve our framework for in situ deployments and large-scale time-varying data.
We would also like to mitigate the effects of sequential topological simplifications. 
It may be possible to improve the computation efficiency by reducing the number of iterations to eliminate false cases using machine learning techniques. 
Finally, we are interested in preserving topological features using gradient-based topological descriptors, such as Morse and Morse--Smale complexes.

\acknowledgments{
This research was partially supported by DOE DE-SC0023157, DOE DE-SC0022753, DOE DE-SC0021015, NSF IIS 1910733, NSF IIS 2145499, NSF OAC-2311878, and NSF OAC-2330367. 
}

\clearpage
\appendix
\section{Details on Experimental Datasets}
\label{sec:data-details}

The $\HF$ dataset comes from the simulation of a 2D flow generated by a heated cylinder using the Boussinesq approximation~\cite{GuntherGrossTheisel2017, Popinet2004}. 
We convert one time instance of the flow into a scalar field using the magnitude of the velocity vector. The dataset is available via the Computer Graphics Laboratory~\cite{CGL}.

The $\EW$ dataset is a 2D scalar field processed using a HiResMIP-v1.0 (1950-Control) dataset~\cite{CaldwellMametjanovTang2019} from the Energy Exascale Earth System Model (E3SM)~\cite{GolazCaldwellVan2019} project~\cite{E3SM}. 
We use the magnitude of \emph{UBOT} and \emph{VBOT} parameters as scalar fields, which correspond to the lowest model level zonal and meridional wind, respectively. 

The $\VF$ dataset is a snapshot of a simulation run capturing the viscous fingers, that is, areas of high concentration during diffusion. 
During the simulation, a cylinder is filled with water with an unlimited supply of salt at the top of the cylinder. The simulation captures the diffusion of the salt as higher density salt solution sinks down in the cylinder. 
This dataset originates from the IEEE Scientific Visualization Contest 2016~\cite{SVC2016}. 
 
The $\TN$ dataset is a 3D synthetic model of a tornado created by Roger Crawfis~\cite{Tornado2003}. 
The flow is scaled to a larger domain and sampled onto a regular grid. 
It is also available via~\cite{CGL}.

The $\AF$ dataset contains one instance from the simulation of an incompressible 3D flow around a CAD model of the Research Vessel Tangaroa~\cite{PopinetSmithStevens2004}. 
We use the magnitude of the velocity vector as the 3D scalar field.

The $\IB$ dataset originates from the IEEE Scientific Visualization Contest 2004~\cite{SVC2004}. It is a simulation of a hurricane from the National Center for Atmospheric Research in the United States. We use the wind speed field and truncate $500 \times 500 \times 90$ from the original $500 \times 500 \times 100$ volume to avoid ``no data'' values on land.

\myedit{The $\NYX$ dataset in~\cref{sec:nyx} is a post analysis cosmological simulation dataset composed of 3D arrays in space~\cite{Biwer2019}. It is based on the Lawrence Berkeley National Laboratory (LBNL) compressible cosmological hydrodynamics simulation code \emph{Nyx}~\cite{AlmgrenBellLijewski2013} that solves equations of compressible hydrodynamics flows in an expanding universe. We use dark matter density as the scalar field with the original $512^3$ volume, together with truncated $128^3$ and $256^3$ volumes for a performance analysis.}

\section{Evaluation Metrics}
\label{sec:metrics}

We review several metrics used for evaluating the compression results. 
\para{Number of false cases.} 
Our method can eliminate all false cases in the decompressed data. 
We report the number of false cases in the decompressed data when comparing {\tool} with off-the-shelf topology-agnostic lossy compressors.

\para{Data compression ratio.}
The data compression ratio is defined to be the ratio between the uncompressed size and compressed size of the input data. 

\para{PSNR.}
Let $f$ and $f'$ denote the original and the decompressed scalar fields.  
The \emph{Peak Signal to Noise Ratio} (PSNR) is defined as 
\begin{align}
PSNR &= 20 \times \log_{10}\left( \frac{\max{(f)}}{\sqrt{MSE}} \right),\\
MSE &= \frac{1}{N} \sum_{i=0}^{N-1} || f(x_i) - f'(x_i)||^2. 
\end{align}
where $||\cdot||$ denotes the $L^2$-norm. 
 
\para{Bottleneck and Wasserstein distances.} 
We use two topology-based metrics to evaluate how much topology is preserved between $f$ and $f'$.
 Let $D$ and $D'$ denote $0$-dimensional persistence diagrams of $f$ and $f'$, respectively.  
Let $\eta$ denote a bijection $\eta: D \to D'$. 
The \emph{bottleneck distance} between $D$ and $D'$ is defined as~\cite{Cohen-SteinerEdelsbrunnerHarer2007}
\begin{align}
W_{\infty}(D, D') =  \adjustlimits \inf_{\eta: D \to D'}  \sup_{p \in D} ||p - \eta(p)||_\infty.
\end{align} 
The \emph{q-Wasserstein distance}~\cite[page 183]{EdelsbrunnerHarer2010} is
\begin{align}
W_{q}(D, D') = \left[ \adjustlimits \inf_{\eta: D \to D'}  \sum_{p \in D} ||p-\eta(p)||^q_\infty \right]^{1/q}.
\end{align} 
We set $q=2$ and quantify the topological differences between $f$ and $f'$ using  
\begin{align}
	d_B(f,f') &= W_\infty(D,D'),\\
	d_W(f,f') &=W_2(D,D').
\end{align}

\section{\myedit{Compression Across Persistence Thresholds}}
\label{sec:sensitivity}

\begin{figure}[!ht]
    \centering
    \includegraphics[width=1.0\columnwidth]{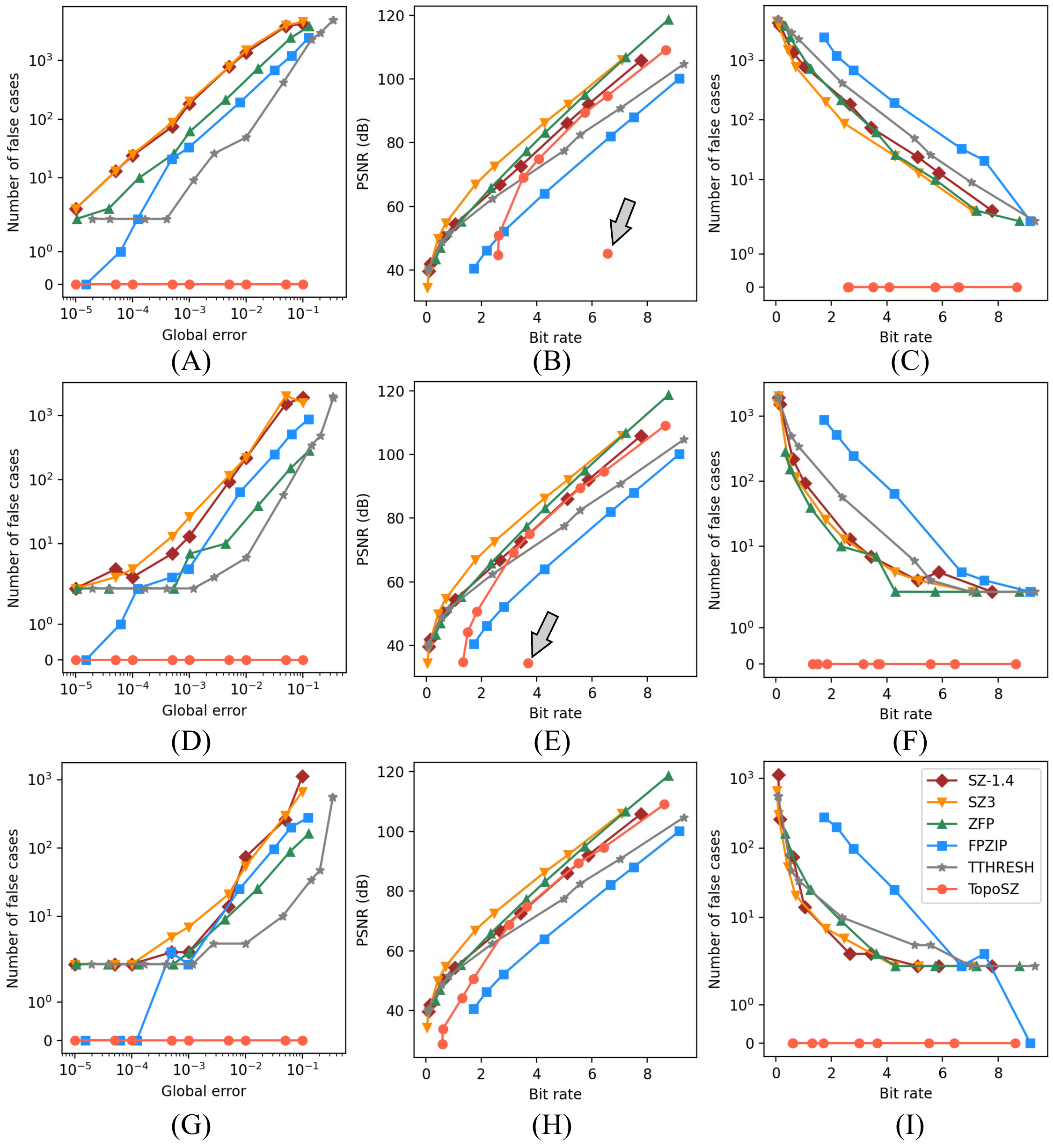}
    \vspace{-6mm}
    \caption{Test lossy compressors with the $\VF$ dataset across three persistence thresholds. From left to right:  the number of false cases \wrt~global error bound; the PSNR; and the number of false cases \wrt~bit rate (i.e.,~average bits per compressed data sample). Lossy compressors include {\tool}, SZ3, ZFP, FPZIP, and TTHRESH. Persistence threshold $\varepsilon=0.02$ (A-C), $\varepsilon=0.06$ (D-F), and $\varepsilon=0.18$ (G-I). All figures use the same color encoding in (I).} 
    \label{fig:rateDistortion2}
\end{figure}

We test various lossy compressors with the $\VF$ dataset across multiple persistence thresholds, in addition to the  results shown in~\cref{fig:false-cases}. 
As shown in~\cref{fig:rateDistortion2}, we reran the experiments in~\cref{sec:result-EW} with persistence threshold $\varepsilon=0.02$ (A-C), $\varepsilon=0.06$  (D-F), and $\varepsilon=0.18$ (G-I). 
We obtain the same observations from~\cref{sec:result-EW}. 
First, {\tool} outperforms all the error-bounded compressors on preserving topology. 
Second, {\tool} has a slightly worse rate distortion in terms of PSNR, compared to SZ3 and ZFP.

We could use any persistence threshold with {\tool}. 
In practice, we recommend setting $\xi >\varepsilon$, since a smaller global error bound $\xi$ typically leads to smaller topological regions that need fine-grained controls. This would lead to less iterations and a higher compression ratio. 

{\tool} produces results with low compression ratios when we set $\xi=0.1$ and $\varepsilon=0.02$, as pointed by an arrow in ~\cref{fig:rateDistortion2} (B). 
{\tool} performs similarly at $\xi=0.1$ and $\varepsilon=0.06$, see an arrow in~\cref{fig:rateDistortion2} (E).
This phenomenon happens because a larger global error bound $\xi$ is more lenient toward false cases, whose persistence are larger than $\vareps$ and less than $\xi$.
These false cases require finer control thus more iterations, whereas more interactions decrease the compression ratio.

\begin{figure*}[t]
    \centering
    \includegraphics[width=1.0\textwidth]{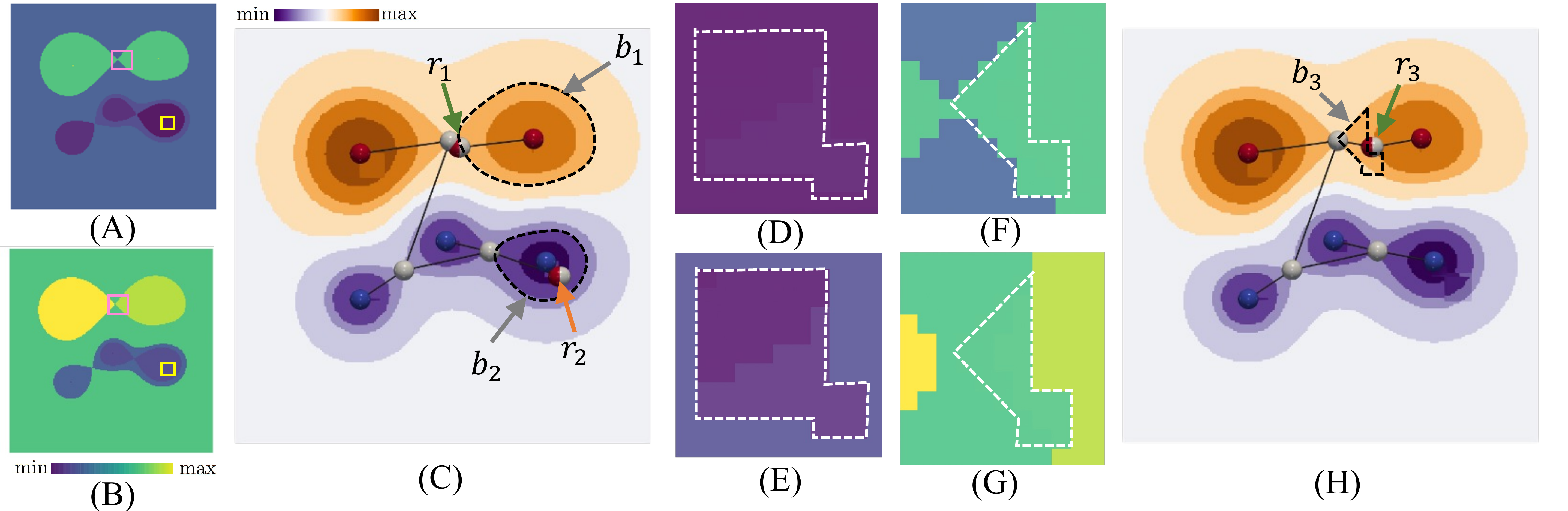}
    \vspace{-6mm}
    \caption{Lower (A) and upper (B) bounds of {\tool} in the  initialization step for the dataset in~\cref{fig:ct-segmentation}. (C) shows the decompressed scalar field after initialization. (D-E): Parts of the updated lower (D) and upper (E) bounds to eliminate false positive case $r_2$. (D) and (E) are zoomed-in views of the  yellow boxes in (A) and (B), respectively. (F-G): Parts of updated lower (F) and upper (G) bounds to eliminate the false positive case $r_1$. (F) and (G) are zoomed-in views of pink boxes in (A) and (B), respectively. (H) shows the decompressed scalar field after the 1st iteration.} 
    \label{fig:FP1}
    \vspace{-2mm}
\end{figure*}
\section{\myedit{Additional Experiments}}
\label{sec:nyx}

\begin{figure}[!ht]
    \centering
    \includegraphics[width=1.0\columnwidth]{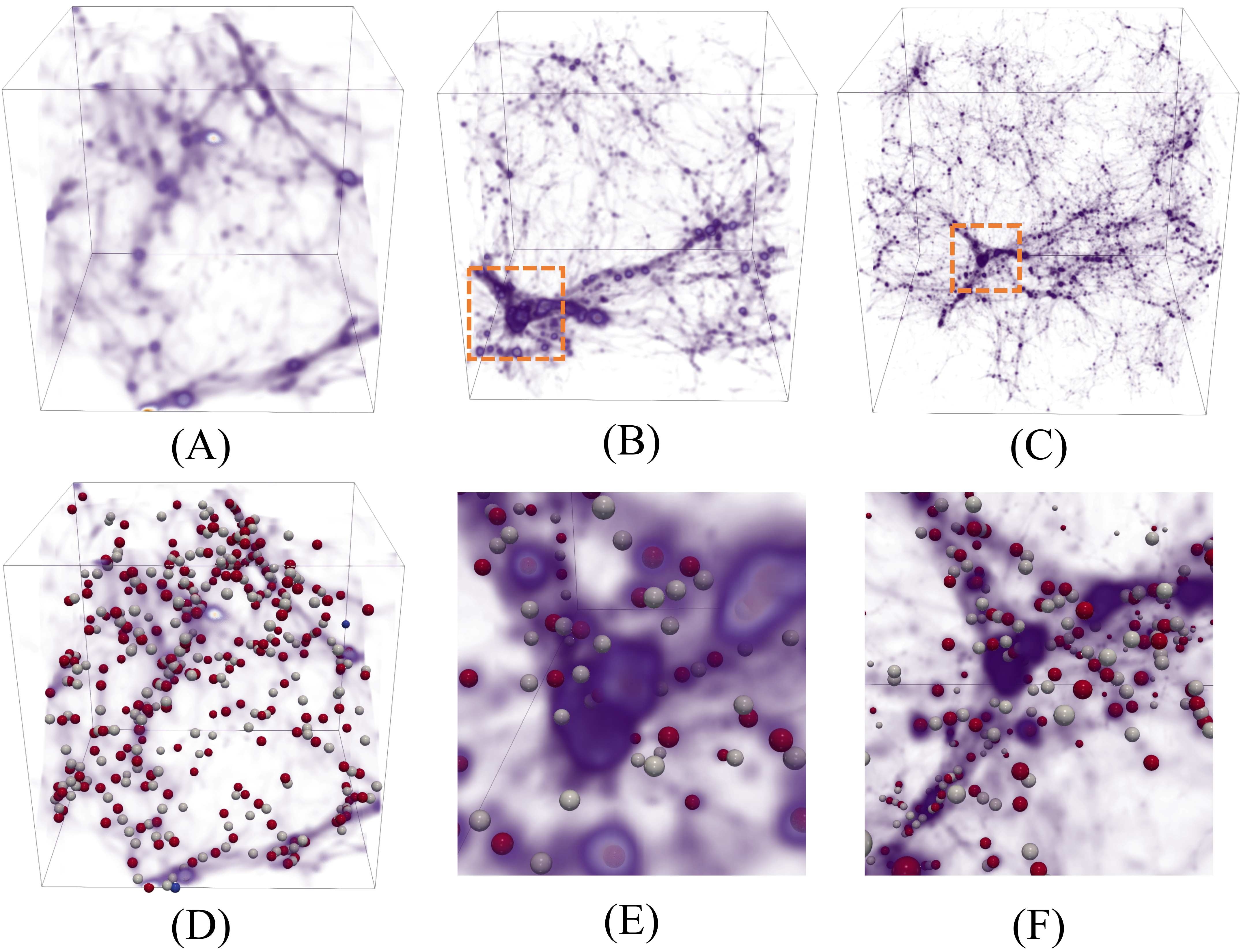}
    \vspace{-6mm}
    \caption{3D visualization of the $\NYX$ dataset with two truncated volumes (A) $128^3$ and (B) $256^3$, and the original volume  (C) $512^3$. (D) is (A) overlaid with critical points. (E) and (F) are the zoomed-in views of the orange boxes in (B) and (C), respectively, overlaid with critical points.} 
    \label{fig:nyx}
\end{figure}

We perform additional experiments with the $\NYX$ dataset. 
We use the $\NYX$ dataset with the original ($512^3$) and two truncated  volumes ($128^3$ and $256^3$), respectively, as shown in~\cref{fig:nyx}(A-C). 
We investigate the relationship between the compression quality, the  run time, and the size of data.  
Each truncated volume is a subset of the original volume with high feature density; see critical points in~\cref{fig:nyx}(D-F). 
Table~\ref{table:nyx} provides run time and compression quality (PSNR and compression ratio) of {\tool} with a persistence threshold $\vareps=0.01$ and a global error bound $\xi=0.005$. 
We observe that the larger volume needs more iterations to eliminate all false cases, and more run time for each iteration and the initialization. 
The PSNR does not change much with the increasing data size, whereas the compression ratio increases with the data size under a uniform parameter setting.

\begin{table}[!ht]
\caption{Run time and performance analysis of {\tool} using the $\NYX$ dataset with varying dimensions. Dim., $\#$CP, and RO represent the dimension (the size of each volume), the number of critical points, and the compression ratio, respectively. Other notations follow the Table~\ref{table:all-time}. All times are in seconds.} 
\label{table:nyx}
\centering
\resizebox{\columnwidth}{!}{
\begin{tabular}{ccccccccccc}
\toprule
   \multirow{2}{*}{\textbf{Dim.}} & \multicolumn{3}{c}{\textbf{Initialization}} & \multicolumn{4}{c}{\textbf{Iteration}} &
   \multirow{2}{*}{\textbf{$\#$CP}}
   & \multirow{2}{*}{\textbf{PSNR}}
   & \multirow{2}{*}{\textbf{RO}}\\
  \cmidrule(lr){2-4}\cmidrule(lr){5-8}
   &\textbf{CT}&\textbf{UBLB} & \textbf{CSZ-1.4} &\textbf{CT}&\textbf{UBLB} & \textbf{CSZ-1.4}&\textbf{\#}&&&\\
\midrule
	$128^3$ & 27.64 & 2.59 & 0.20  & - & - & - & 0 & 400&76.54&16.7\\
	$256^3$ & 144.89 & 24.02 & 1.19 & 90.25 & 40.27 & 1.13 & 3& 444&71.8&63.9\\
	$512^3$ &2425.99 & 200.31 & 11.18 & 826.16 & 806.74 & 10.24 & 5 & 2,474&72.6&78.8\\
\bottomrule
\end{tabular}
}
\end{table}
\section{\myedit{A Worst-Case Analysis of TopoSZ}}
\label{sec:worst-case}

Given the nature of the expanding $k$-layers (see~\cref{sec:iteration}), it may be possible (in a worst-case scenario) that $k$ expands to the entire dataset, forcing lossless encoding of a significant fraction of the data. 
Theoretically, {\tool} does not formally guarantee that such a worst-case scenario would not occur. 
However, we can study when iterations and the expansion of $k$-layer neighborhoods are needed, using the example dataset from~\cref{fig:ct-segmentation}.
To force {\tool} to run multiple iterations, we set the persistence threshold $\vareps$ and the global error bound $\xi$ to be relatively high, that is, $\varepsilon=0.1$ and $\xi=0.2$, respectively.

\cref{fig:FP1} (A) and (B) visualize the lower and upper bounds during the initialization step. 
\cref{fig:FP1} (C) shows that there are two false positive cases, marked as critical points $r_1$ and $r_2$ after the initialization. 
$r_2$ is located within a topological region (\ie, a contour-tree-induced segment) that is  bounded by the curve $b_2$ in \cref{fig:FP1} (C). 
False cases such as $r_2$ are easy to eliminate and usually disappear when we give finer-grained bounds for their corresponding regions. 
Indeed, $r_2$ disappears after the 1st iteration, when we update the lower and upper bounds within regions \cref{fig:FP1} (D) and (E), which are zoomed views of the regions bounded by yellow boxes in \cref{fig:FP1} (A) and (B), respectively.

False cases located at the boundary of topological regions are harder to eliminate. 
They are usually the reason behind multiple iterations. 
Such false cases occur because each topological region has its own lower and upper bounds and, therefore, its own ``local'' compression and decompression process. 
When two decompressed topological regions are ``glued'' with each other, some false cases may occur on their shared  boundary. 
These false cases are harder to eliminate because a new iteration might create a new, but smaller, topological region and false cases may occur on the its boundary. 

For example, $r_1$ in~\cref{fig:FP1} (C) is on the boundary of a topological region bounded by the curve $b_1$. 
In order to eliminate $r_1$, {\tool} runs the 1st iteration with an updated lower and upper bounds (\cref{fig:FP1} (F) and (G)). 
However, a new critical point $r_3$ in~\cref{fig:FP1} (H) appears on this new boundary (bounded by the curve $b_3$) after the 1st iteration. 

In practice, the good news is that false cases that appear on the boundary of topological regions are usually located in tiny regions of the domain and correspond to tiny branches of the contour tree. 
Therefore, we only need to update the lower and upper bounds of a tiny region in the domain to eliminate them. 
We also observe that in practice, these cases may disappear and appear pixel by pixel (or voxel by voxel) during iterations, if we use a 1-layer neighborhood expansion. 
Therefore, we use expanding $k$-layer neighborhood to reduce the number of iterations while sacrificing some compression ratio. 
Since these false cases occur in tiny regions of the domain, they are easy to eliminate with a small number of iterations. 
For example, in the experiment of~\cref{fig:FP1}, all false cases are eliminated in the 2nd iteration. 
Across all our experiments, we never encountered the worst-case scenario. 

Finally, since the performance of {\tool} (in terms of topology preservation) improves during iterations (see~\cref{sec:results-iterations}), we may terminate the compression process after a fixed number of iterations or arriving at a fixed number of false cases, to tolerate the rare worst-case scenario. 

For example, when we run {\tool} with the $\NYX$ dataset, the number of false cases after initialization is around 400, which decreases to 4 after 8 iterations. 
In this case, if {\tool} is terminated after 8 iterations, almost all topological information is preserved with less compression time and a larger compression ratio, compared with removing all false cases.

\section{\myedit{Pointwise Error Control of TopoQZ and TopoSZ}}
\label{sec:error-control}
\begin{figure}[!ht]
    \centering
    \includegraphics[width=1.0\columnwidth]{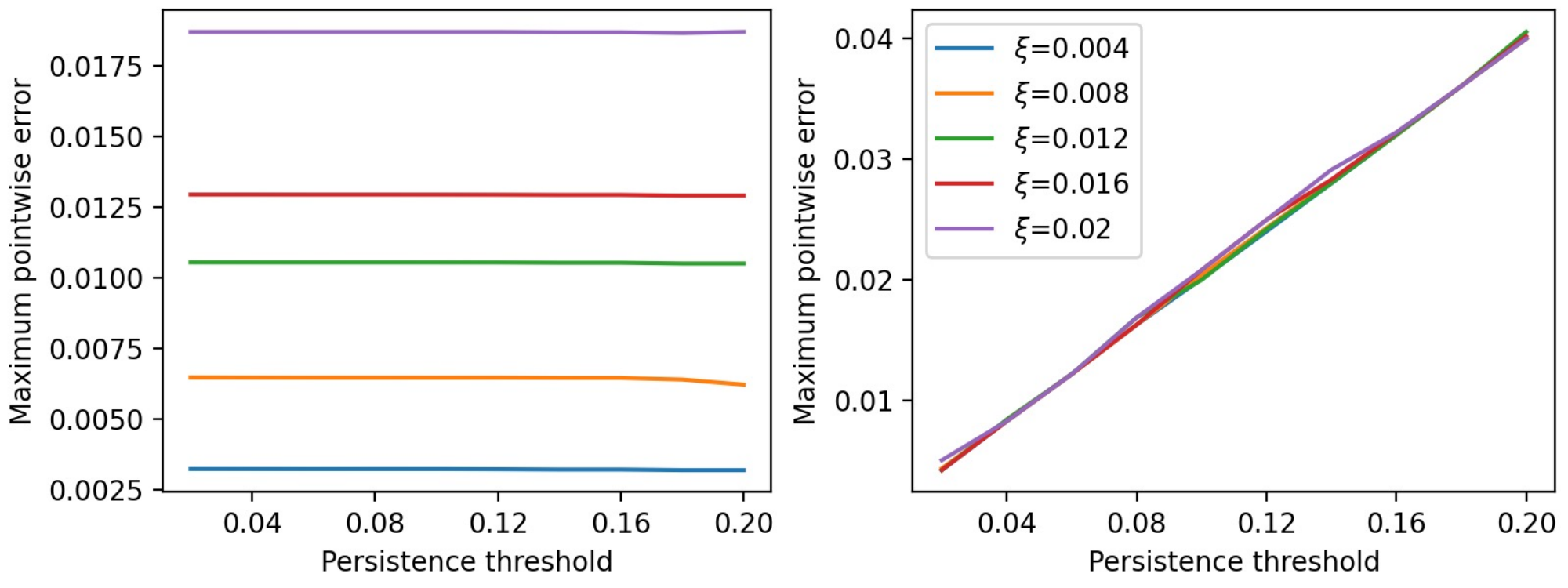}
    \vspace{-6mm}
    \caption{The average maximum pointwise difference between the original and the decompressed data  using {\tool} (left) and TopoQZ (right).} 
    \label{fig:pointwise-control}
\end{figure}

\cref{fig:pointwise-control} demonstrates the evolution of the maximum pointwise error between the original and the decompressed data, averaged over all  datasets in Table~\ref{table:data}, for an increasing persistence threshold with different global error bounds $\xi$. 
\cref{fig:pointwise-control} (left) indicates that {\tool} has a strict control on pointwise error, whereas TopoQZ does not, as shown in~\cref{fig:pointwise-control} (right). 
Therefore, to the best of our knowledge, {\tool} is the first lossy compressor that combines pointwise error control and topological guarantee during compression. 

\end{document}